\documentclass[conference]{IEEEtran} 
\IEEEoverridecommandlockouts
\usepackage{cite}
\usepackage{amsmath,amssymb,amsfonts}
\usepackage{booktabs}
\usepackage{algorithmic}
\usepackage{graphicx}
\usepackage{textcomp}
\def\BibTeX{{\rm B\kern-.05em{\sc i\kern-.025em b}\kern-.08em
    T\kern-.1667em\lower.7ex\hbox{E}\kern-.125emX}}
\usepackage[caption=false]{subfig}
\usepackage{xspace}
\usepackage{multirow}
\usepackage{array}
\usepackage[dvipsnames,table,xcdraw]{xcolor}
\usepackage{nohyperref}  
\usepackage{url}  
\usepackage{tikz}
\usetikzlibrary{positioning, intersections, arrows, shapes.geometric, calc,
  shapes.callouts, decorations.pathreplacing, decorations.markings, shapes, patterns}
\usepackage{pifont}
\usepackage{pgfplots}
\usepgfplotslibrary{statistics}
\usepackage{pgfplotstable}
\pgfplotsset{compat=newest}
\usepackage{filecontents}
\usepackage[utf8x]{inputenc}
\usepackage{lipsum}
\usepackage{soul}
\usepackage{comment}
\usepackage{amsthm}
\usepackage{mdframed}
\usepackage[most]{tcolorbox}
\usepackage{blindtext}
\usepackage{color,soul}

{\endMakeFramed}

\newenvironment{frshaded*}{%
\MakeFramed {\advance\hsize-\width \FrameRestore}}%
{\endMakeFramed}

\graphicspath{{figure/}{figures/}}

\AtBeginDocument{%
  \providecommand\BibTeX{{%
    \normalfont B\kern-0.5em{\scshape i\kern-0.25em b}\kern-0.8em\TeX}}}

\newcommand{\datanum}{300,000\xspace}

\newcommand\FIG{Fig.\xspace}

\skip\footins 0.21cm
\makeatletter
\def\footnoterule{\relax%
  \kern0pt
  \hbox to \columnwidth{\hfill\vrule width 1.0\columnwidth height 0.4pt\hfill}
  \kern4.6pt}
\makeatother

\begin{document}

\title{\huge An Empirical Analysis of VM Startup Times in Public IaaS Clouds: An Extended Report}

\author{\IEEEauthorblockN{Jianwei Hao\IEEEauthorrefmark{2}\IEEEauthorrefmark{1}\thanks{* Both authors contributed equally.},
    Ting Jiang\IEEEauthorrefmark{2}\IEEEauthorrefmark{1},
    Wei Wang\IEEEauthorrefmark{3},
    and In Kee Kim\IEEEauthorrefmark{2}}
    \IEEEauthorblockA{\IEEEauthorrefmark{2}University of Georgia, Department of Computer Science, \{jhao, ting.jiang1, inkee.kim\}@uga.edu\\
        \IEEEauthorrefmark{3}The University of Texas, San Antonio, Department of Computer Science, wei.wang@utsa.edu
    }
}

\maketitle

\thispagestyle{plain}
\pagestyle{plain}

\begin{abstract}
VM startup time is an essential factor in designing elastic cloud applications. For example, a cloud application with autoscaling can reduce under- and over-provisioning of VM instances with a precise estimation of VM startup time, and in turn, it is likely to guarantee the application's performance and improve the cost efficiency. However, VM startup time has been little studied, and available measurement results performed previously did not consider various configurations of VMs for modern cloud applications. 

In this work, we perform comprehensive measurements and analysis of VM startup time from two major cloud providers, namely Amazon Web Services (AWS) and Google Cloud Platform (GCP). With three months of measurements, we collected more than \datanum data points from each provider by applying a set of configurations, including 11+ VM types, four different data center locations, four VM image sizes, two OS types, and two purchase models (e.g., spot/preemptible VMs vs. on-demand VMs). With extensive analysis, we found that VM startup time can vary significantly because of several important factors, such as VM image sizes, data center locations, VM types, and OS types.  Moreover, by comparing with previous measurement results, we confirm that cloud providers (specifically AWS) made significant improvements for the VM startup times and currently have much quicker VM startup times than in the past.

\end{abstract}


\section{Introduction}\label{sec:introduction}
For the last decade, cloud computing has become a primary computing infrastructure as many applications have been increasingly migrated from on-premise environments to clouds~\cite{ViewCloud:CACM10, AWS-online, GoogleCloud-online, OpenStack-online}. At the same time, cloud infrastructure itself has been continuously evolving so that cloud computing currently offers diverse resource models, such as VMs, containers~\cite{Docker-online, gvisor-online} and orchestration~\cite{MESOS:NSDI11, Borg:Eurosys20}, and cloud functions~\cite{AWSLambda-online, Occupy:SoCC17, CACM19:RiseServerless}, to support various application types and service scenarios.
While VMs are the most traditional resource type in the clouds, VMs are still widely used as common hosting platforms for both user applications and different resource models, like containers~\cite{ecs-online}, Kubernetes~\cite{k8s-online}, and serverless/cloud functions~\cite{Catalyzer:ASPLOS2020}.

Various aspects of performance implications in VMs and IaaS (Infrastructure as a Service) clouds, such as Amazon EC2, have been extensively studied~\cite{RuntimeMeasurement:VLDB10, PerfVar:TPDS11, Chaos:TIT2016, PerfVarWei:FSE19, EarlyObservation:HPDC10, MemPerf:TCC17, IOPerf:JGC13, NetPerf:CN15, Less:CLOUD15, AWSNetwork:Sigmetrics15, StartupMao:CLOUD12}. The measurement results are widely adopted for developing novel cloud applications and cloud infrastructure management systems.
In particular, understanding VM startup time~\cite{Provisioning:CCGrid14, StartupMao:CLOUD12, Provisioning:Bench18} is crucial to design elastic resource management systems for cloud applications, such as autoscaling
~\cite{Ming-Autscale-SC11,PredEval:Cloud2016}.
According to a previous study~\cite{StartupMao:CLOUD12} performed in 2012, VM startup time could vary significantly due to various factors, such as VM image sizes, time-of-day, VM purchase models (on-demand or spot), etc. While the measurement results from the previous work are still useful to current cloud research, a number of mechanisms have been proposed for optimizing VM startup processes in cloud data centers for the last decade~\cite{ScalabeVMDeployment:SC13, Squirrel:HPDC14, muVM:ICDCS15, Xu:JSS16, LigherVM-OSDI2017, CooporativeCaching:TCC2018, YOLO:Europar2019, Firecracker-NSDI2020}. Therefore, it is important to see whether there is an improvement of VM startup time in public cloud providers.

This work aims to provide {\em up-to-date} information about VM startup time to the research community so that cloud researchers and practitioners facilitate the design of novel resource and performance management approaches for cloud applications.
To this end, we performed an empirical analysis of VM startup times, measured from two widely-used public cloud services, namely AWS (Amazon Web Services)~\cite{AWS-online} and GCP (Google Cloud Platform)~\cite{GoogleCloud-online}. The measurement was conducted for three months with extensive trials. Each trial was performed for at least 14 consecutive days to cover the temporal impact (time-of-day, date-of-week)~\cite{PerfVar:TPDS11, TPDS11:Iosup, PerfVarWei:FSE19} on VM startup time. In total, we obtained more than \datanum data points from each provider by exploiting a different set of factors that can change the VM startup time. The data was collected by using 11+ widely-used VM types, which are commonly hosted in four service regions, located in the U.S., Europe, and Asia. Four VM image sizes (from 32GB to 256GB), two different OS types (Linux and Windows), and two VM purchase models (on-demand and spot/preemptible) were also applied. 
Moreover, we measured two different types of VM startup time; a {\em cold startup} and {\em warm startup} time. The cold startup time means the VM's startup time when a user creates a new VM so that this startup time is equivalent to the VM's creation (provisioning) time. The warm startup time is the startup time measured when a user (re)starts an existing (and stopped) VM instance.

This paper reports (cold/warm) VM startup times of AWS and GCP with diverse configurations for realistic VM use-cases. With extensive analysis, we found that VMs' startup time could be significantly changed due to several factors, such as OS types, VM image sizes, VM types (and generation), and locations/regions. 
Table~\ref{tab:analysis_results} summarizes our analysis and important factors for changing VM startup times.

\begin{table*}[t]
\centering
\caption{Important Factors and Analysis Results of VM Startup Times.}
\begin{tabular}{|c|c|l|c|}
\hline
{\bf Category} & {\bf Provider} & {\bf Observation} & {\bf Proof} \\ \hline
\multirow{3}{*}{\begin{tabular}[c]{@{}c@{}}OS\\ Types\end{tabular}} & Both & \begin{tabular}[c]{@{}l@{}}{\bf AWS VMs have faster startup times than VMs in GCP.} AWS VMs showed 2.22× (Linux VMs) and \\ 1.2× (Windows VMs) shorter cold startup times compared to VMs in GCP.\end{tabular} & Table~\ref{tab:all_startup} \\ \cline{2-4} 
 & AWS & {\bf Linux VMs have a 38\% shorter cold startup time} than Windows VMs. & Table~\ref{tab:all_startup} \\ \cline{2-4} 
 & GCP & {\bf Windows VMs have 15\% -- 46\% shorter startup times} compared to Linux VMs. & Table~\ref{tab:all_startup} \\ \hline
\multirow{3}{*}{\begin{tabular}[c]{@{}c@{}}Image\\ Size\end{tabular}} & AWS & \begin{tabular}[c]{@{}l@{}}{\bf AWS VMs' startup times are independent of the VM image size}, showing that VMs with different \\ image sizes often have similar startup time.\end{tabular} & \FIG~\ref{fig:aws_vm_size} \\ \cline{2-4} 
 & \multirow{2}{*}{GCP} & \begin{tabular}[c]{@{}l@{}}{\bf VM's startup times have a positive correlation with the VM image size.} i.e., Startup times of VMs\\ with 128GB (of the image size) are longer than that of VMs with 64GB.\end{tabular} & \FIG~\ref{fig:gcp_vm_size} \\ \cline{3-4} 
 &  & \begin{tabular}[c]{@{}l@{}} {\bf A caching-based mechanism is used to boost VM startup process} by temporarily storing recently used \\ VM images. VM startup times. This mechanism can (sometimes) diminish the impact of VM image sizes.\end{tabular} & \begin{tabular}[c]{@{}l@{}} \FIG~\ref{fig:google_cache} and \\ Table~\ref{tab:cache_period} \end{tabular}  \\ \hline
\multirow{4}{*}{\begin{tabular}[c]{@{}c@{}}Instance\\ Type\end{tabular}} & Both & {\bf Different VM types have different startup times.} & \FIG~\ref{fig:aws_vmtype} and \ref{fig:google_vmtype_linux} \\ \cline{2-4} 
 & AWS & {\bf The older generation VM types have longer startup times} than the newer generation VM types. & \FIG~\ref{fig:aws_vmtype_gen} \\ \cline{2-4} 
 & \multirow{2}{*}{GCP} & {\bf The older generation Linux VMs have longer startup times} than the newer generation Linux VMs. & \FIG~\ref{fig:google_vmtype_gen} \\ \cline{3-4} 
 &  & \begin{tabular}[c]{@{}l@{}} {Windows VMs can be the opposite.} {\bf Newer (e.g., 2\textsuperscript{nd}) generation VMs can have longer startup times } \\ {\bf than older (e.g., 1\textsuperscript{st}) generation VMs.} \end{tabular} & \FIG~\ref{fig:google_vmtype_gen} \\ \hline
\multirow{2}{*}{Location} & AWS & \begin{tabular}[c]{@{}l@{}} On average, VMs in different regions have similar startup times. But, {\bf VMs in different zones in each} \\ {\bf region have substantially different startup times (up to 45\%).}\end{tabular} & \FIG~\ref{fig:byregion} and \ref{fig:byzoneAWS} \\ \cline{2-4} 
 & GCP & \begin{tabular}[c]{@{}l@{}} {\bf VMs in different regions show distinct VM startup times.} Moreover, {\bf VMs in different zones often } \\ \bf{show significant changes (up to 50\%) in VM startup times within the same region.}\end{tabular} & \FIG~\ref{fig:byregion} and \ref{fig:byzoneGoogle} \\ \hline
 \begin{tabular}[c]{@{}c@{}}Purchase\\ Model\end{tabular} & Both & \begin{tabular}[c]{@{}l@{}} Spot (AWS) and preemptible (GCP) VMs have similar startup time compared to on-demand VMs, \\ indicating that {\bf spot or preemptible VMs no longer have slower startup times than on-demand VMs.}\end{tabular} & \begin{tabular}[c]{@{}l@{}} \FIG~\ref{fig:ODSpotAWS} and \\ Table~\ref{tab:google_preemptive} \end{tabular} \\ \hline
\end{tabular}
\label{tab:analysis_results}
\end{table*}

After identifying the critical factors for VM startup times, we compare our analysis results with the previous studies~\cite{StartupMao:CLOUD12, Provisioning:Bench18}. The comparison confirms that cloud providers (specifically AWS) made significant improvements for the VM startup times and currently have much quicker VM startup times. Moreover, implications and findings from this study will help various research in cloud resource and application management. In particular, autoscaling algorithms~\cite{Ming-Autscale-SC11, UCC14:LCA, PredEval:Cloud2016, TOMPECS18:AutoscalingWF, CloudInsight:Cloud2018, IPDPS20:Dynamo, TCC20:CloudInsight} with the accurate VM startup time can determine the exact scaling point for handling increased user demands. And, cloud simulators~\cite{CloudSim:SPE2011, JGC12:iCanCloud, JSC12:GreenCloud, PICS:CLOUD2015, TrustSim-IC2E15} can generate more reliable simulation results with this study.

It is worth noting that this paper is an extended version of our previous work~\cite{CLOUD2021:VMStartup}, which is published in the 2021 IEEE International Conference on Cloud Computing (IEEE CLOUD). This version of the paper contains additional measurements and analysis on VM start times in AWS and GCP, which are not included in the IEEE CLOUD 2021 version.

As a result, this work has the following contributions.

1. We performed a thorough measurement study on VM startup time of two major cloud providers; AWS and GCP, which are widely used in research and industry.

2. With three months of measurement, we collected a large number of data points with various VM configurations, reflecting the realistic use-cases of VMs in clouds. 

3. We report VM startup times, in AWS and GCP, with diverse configurations. With extensive analysis, we found several factors that considerably change the VM startup times.

4. We found that GCP uses a cache-based approach to reduce VMs' startup time. Recently used VM images are stored in the GCP data center for the next 75 -- 100 minutes, and users can benefit from using cached VM images to reduce VM startup times. 

We structure the rest of the paper as follows.
Section~\ref{sec:methdology} describes the experimental setup for measuring VM startup time.
Section~\ref{sec:results} reports the measurement results of VM startup times from both AWS and GCP.
Section~\ref{sec:discussion} provides a comparison with previous works.
Section~\ref{sec:relatedwork} summarizes related work. Finally, Section~\ref{sec:conclusion} concludes this paper.

\section{Measurement Methodology}\label{sec:methdology}
This section describes the methodology for measuring VM startup time from two public cloud providers.

\subsection{Measurement Setup}

We considered diverse factors that can lead to changing the VM startup time. The following configurations were used for this measurement to collect VM startup times with more realistic scenarios.

\vspace{1mm}
\noindent
{\bf Cloud Providers.} We measured VM startup time from AWS~\cite{AWS-online} and GCP~\cite{GoogleCloud-online}. Because these two providers are widely used in both industry and academic research, it is crucial to see if there VM startup time differences exist in both providers. We used VMs from AWS EC2 and Google Compute Engine, which are IaaS models of these providers. 

\vspace{1mm}
\noindent
{\bf Measurement Period.} The measurement was conducted for three months in 2020 with a set of trials. Each measurement trial has at least two-weeks of duration to check there are temporal impacts on the VM startup time, such as time-of-day, day-of-week.

\vspace{1mm}
\noindent
{\bf Data Center Locations (Regions and Availability Zones).} Table~\ref{tab:regions} describes the regions and (availability) zones of two providers used for this measurement. We chose four regions, located in the U.S. (east and west), Europe, and Asia, to check any correlations between cloud data center locations and the VM startup times. Moreover, each region has multiple zones so that we measured VM startup time from all the zones listed in Table~\ref{tab:regions} to see if there are VM startup time fluctuations with different 
(availability) 
zones in the regions. 

\begin{table}[t]
\centering
\caption{Data center regions and (availability) zones.}
\begin{tabular}{|l|l|l|l|}
\hline
{\bf Provider}          & {\bf Region}    & {\bf Zones} & {\bf Location}    \\ \hline
\multirow{4}{*}{AWS}    & \texttt{us-east-1}       & 5 Zones (a -- d, f)   & N. Virginia \\ \cline{2-4} 
                        & \texttt{us-west-2}       & 3 Zones (a -- c)   & Oregon      \\ \cline{2-4} 
                        & \texttt{eu-west-3}       & 3 Zones (a -- c)   & Paris       \\ \cline{2-4} 
                        & \texttt{ap-southeast-1}  & 3 Zones (a -- c)   & Singapore   \\ \hline
\multirow{4}{*}{GCP} & \texttt{us-east4}        & 3 Zones (a -- c)   & N. Virginia \\ \cline{2-4} 
                        & \texttt{us-west1}        & 3 Zones (a -- c)   & Oregon      \\ \cline{2-4} 
                        & \texttt{europe-west1}    & 3 Zones (b -- d)   & Belgium     \\ \cline{2-4} 
                        & \texttt{asia-southeast1} & 3 Zones (a -- c)   & Singapore   \\ \hline
\end{tabular}
\label{tab:regions}
\end{table}

\vspace{1mm}
\noindent
{\bf Instance Types.} Table~\ref{tab:vmtypes} shows 12 VM types from AWS and 11 VM types from GCP, used in the measurement to check the VM startup time differences in various instance types. We categorized these VM types into five VM classes: tiny, small, medium, large, and xlarge classes. This classification is based on the memory size of VM types. VM types in each class are not exactly the same, but almost equivalent VM types were chosen from both cloud providers. Moreover, when selecting instance types for the measurement, we considered VMs with different generations and CPU models. For example, we measured the startup time from both {\tt t2.small} and {\tt t3.small} instances (in the Small class) from AWS to see if there is the VM startup time difference from the equivalent VM instances with different generations. For the diversity of CPU models, we used {\tt t3a.medium} and {\tt m5a.large} with AMD CPUs from AWS. {\tt n2} (Intel), {\tt n2d} (AMD), and {\tt e2} (Intel or AMD) instance types from GCP were used for the measurement to check the VM startup time differences due to different CPU models.

\vspace{1mm}
\noindent
{\bf OS Types and VM Image Sizes.} We tested the VM startup times with two different operating systems (Linux and Windows). We used Ubuntu 18.04 LTS for Linux VMs and Windows Server 2016 for Windows VMs. For the VM image sizes, we used four different sizes of user-created VM images, which are 32GB, 64GB, 128GB, and 256GB. The VM images were fully filled by OS and other binary/data files. 

\begin{table}[t]
\centering
\caption{VM types used for measurement.}
\begin{tabular}{|c|c|c|c|c|}
\hline
{\bf Class}   & {\bf \#vCPU}   & \begin{tabular}[c]{@{}c@{}}{\bf Mem} \\ {\bf (GB)}\end{tabular}  & \begin{tabular}[c]{@{}c@{}}{\bf AWS} \\ {\bf VM Types}\end{tabular} & \begin{tabular}[c]{@{}c@{}}{\bf GCP} \\ {\bf VM Types}\end{tabular} \\ \hline
tiny    & 0.5-1  & 0.5-1    & \begin{tabular}[c]{@{}l@{}}{\tt t2.nano}, \\ {\tt t2.micro},\\ {\tt t3.nano}, \\ {\tt t3.micro}\end{tabular} & {\tt f1-micro} \\ \hline
small   & 1-2    & 1.7-2    & \begin{tabular}[c]{@{}l@{}}{\tt t2.small}, \\ {\tt t3.small}\end{tabular} & {\tt g1-small} \\ \hline
medium  & 1-2    & 3.75-4   & \begin{tabular}[c]{@{}l@{}}{\tt t3.medium}, \\ {\tt t3a.medium}\end{tabular} & {\tt n1-standard-1} \\ \hline
large   & 2      & 7.5-8    & \begin{tabular}[c]{@{}l@{}}{\tt m4.large}, \\ {\tt m5a.large}\end{tabular}  & \begin{tabular}[c]{@{}l@{}}{\tt n1-standard-2}, \\ {\tt n2-standard-2},\\ {\tt n2d-standard-2}, \\ {\tt e2-standard-2}\end{tabular} \\ \hline
xlarge  & 4      & 15-16    & \begin{tabular}[c]{@{}l@{}}{\tt m4.xlarge}, \\ {\tt m5.xlarge}\end{tabular} & \begin{tabular}[c]{@{}l@{}}{\tt n1-standard-4}, \\ {\tt n2-standard-4},\\ {\tt n2d-standard-4}, \\ {\tt e2-standard-4}\end{tabular} \\ \hline
\multicolumn{3}{|c|}{Total} & {\bf 12 Types} & {\bf 11 Types} \\ \hline
\end{tabular}
\label{tab:vmtypes}
\end{table}

\vspace{1mm}
\noindent
{\bf VM Purchase Models.} Both on-demand and low-availability VM models were used for the measurement. The spot instance model~\cite{AWS-Spot-online} (from AWS) and the preemptible instance model~\cite{GoogleCloud-PVM-online} (from GCP) were used for the low availability models. Specifically, it is interesting to measure the VM startup time from two different VM purchase models because spot/preemptible instances can have different provisioning processes or a different level of availabilities compared to the on-demand models.

\subsection{Measurement Procedure}

To measure the VM startup time, we did not rely on the VM status information provided by the cloud providers because the VM status is often inaccurate~\cite{StartupMao:CLOUD12, MonDelay:FGCS2013}. 
Instead, we implemented and deployed applications that collect VM startup times by interacting with both cloud providers. The measurement applications calculated the VM startup time based on the very first successful remote access to the target VM. For example, 
suppose $t_{request}$ is a time to call cloud APIs to start a VM, and $t_{access}$ is the first time to successfully access (e.g., login) the VM via ssh or RDP. Then, a VM's startup time can be calculated by $t_{access} - t_{request}$.

We measured both cold startup (VM provisioning time) and warm startup time (startup time of an existing VM) of VMs. Both startup times of a VM were measured sequentially. During the measurement period, we measured the cold startup time of VMs every hour on the hour, and then, measured the warm startup time after several minutes. For example, a measurement application sent a request for creating a VM to the providers (via {\tt AWS boto3}~\cite{AWS-boto-online} or {\tt Google Cloud APIs}~\cite{GoogleCloud-API-online}) at the top of the hour (e.g., 1 a.m.), and the {\bf cold VM startup time} was measured when the measurement application could successfully access the VM. 
Then, the measurement application stopped the VM. After several minutes (e.g., at 1:10 a.m.), the application sent another request to start the VM and measured the {\bf warm VM startup time} of the VM with successful remote access, and then, the VM was finally terminated. It is worth noting that the VM startup time generally means the cold VM startup time in this work because we found that the cold startup time varies more significantly as per diverse factors, and the warm startup time is fairly consistent in both providers.

\section{Measurement Results and Analysis}\label{sec:results}
\begin{table}[]
\centering
\caption{Cold and warm startup times of Linux and Windows VMs in AWS and GCP}
\begin{tabular}{|c|c|c|c|c|}
\hline
\multirow{2}{*}{\textbf{OS Type}} & \multicolumn{2}{c|}{\textbf{Cold Startup}} & \multicolumn{2}{c|}{\textbf{Warm Startup}} \\ \cline{2-5} 
 & \textbf{AWS} & \textbf{GCP} & \textbf{AWS} & \textbf{GCP} \\ \hline
Linux VM & 55.9s & 124.1s & 34.0s & 32.8s \\ \hline
Windows VM & 89.7s & 107.5s & 24.5s & 22.2s \\ \hline
\end{tabular}
\label{tab:all_startup}
\end{table}

Our measurement collected more than \datanum data points from each provider. These data include both warm and cold startup times measured with a set of different configurations in Section~\ref{sec:methdology}. In total, we measured VM startup time with 768\footnote{768 is calculated by 2 (Linux or Windows) $\times$ 12 (VM types) $\times$ 4 (Image sizes) $\times$ 4 (Regions) $\times$ 2 (On-demand or Spot)} (for AWS) and 704\footnote{704 is calculated by 2 (Linux or Windows) $\times$ 11 (VM types) $\times$ 4 (Image sizes) $\times$ 4 (Regions) $\times$ 2 (On-demand or Preemptible)} (for GCP) different configurations, and the number of collected samples in each configuration has a range from 350 to 1500. For each configuration, we ensured there were enough samples to obtain accurate VM startup measurement results with high confidence using a method proposed by prior work~\cite{Taming:OSDI2018}.

\subsection{OS Types}\label{subsec:os_types}

Different OS types for VMs are widely-recognized factors that can impact VM startup times as per the previous works~\cite{StartupMao:CLOUD12, ScalabeVMDeployment:SC13, muVM:ICDCS15}. 
Therefore, it is important to see if this factor still changes VM startup time. 
To confirm the impact of OS types on VM startup times, we measured the VM startup time using Linux and Windows VMs with four different VM image sizes (from 32GB to 256GB).

\begin{figure}[t]
\centering
	\begin{minipage}[b]{1\columnwidth}
		\includegraphics[width=\columnwidth]{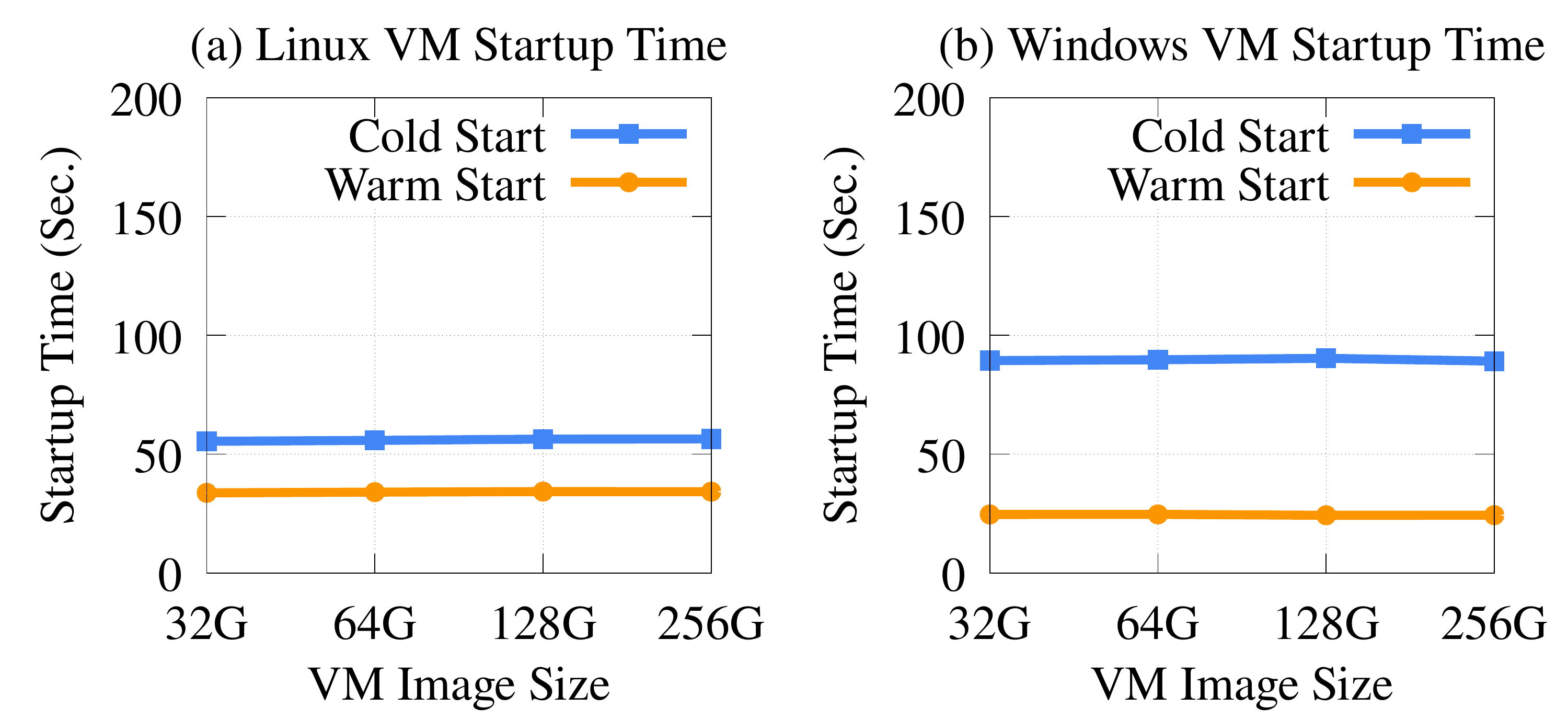}
		\caption{(AWS) VM startup times with different VM image sizes}
		\label{fig:aws_vm_size}
	\end{minipage}

	\vspace{2.00mm}

	\begin{minipage}[b]{1\columnwidth}
	  \subfloat[Cold Startup: Linux VM (64G)\label{fig:aws-vm-cold-l-64g}]{%
	    \includegraphics[width=0.47\columnwidth]{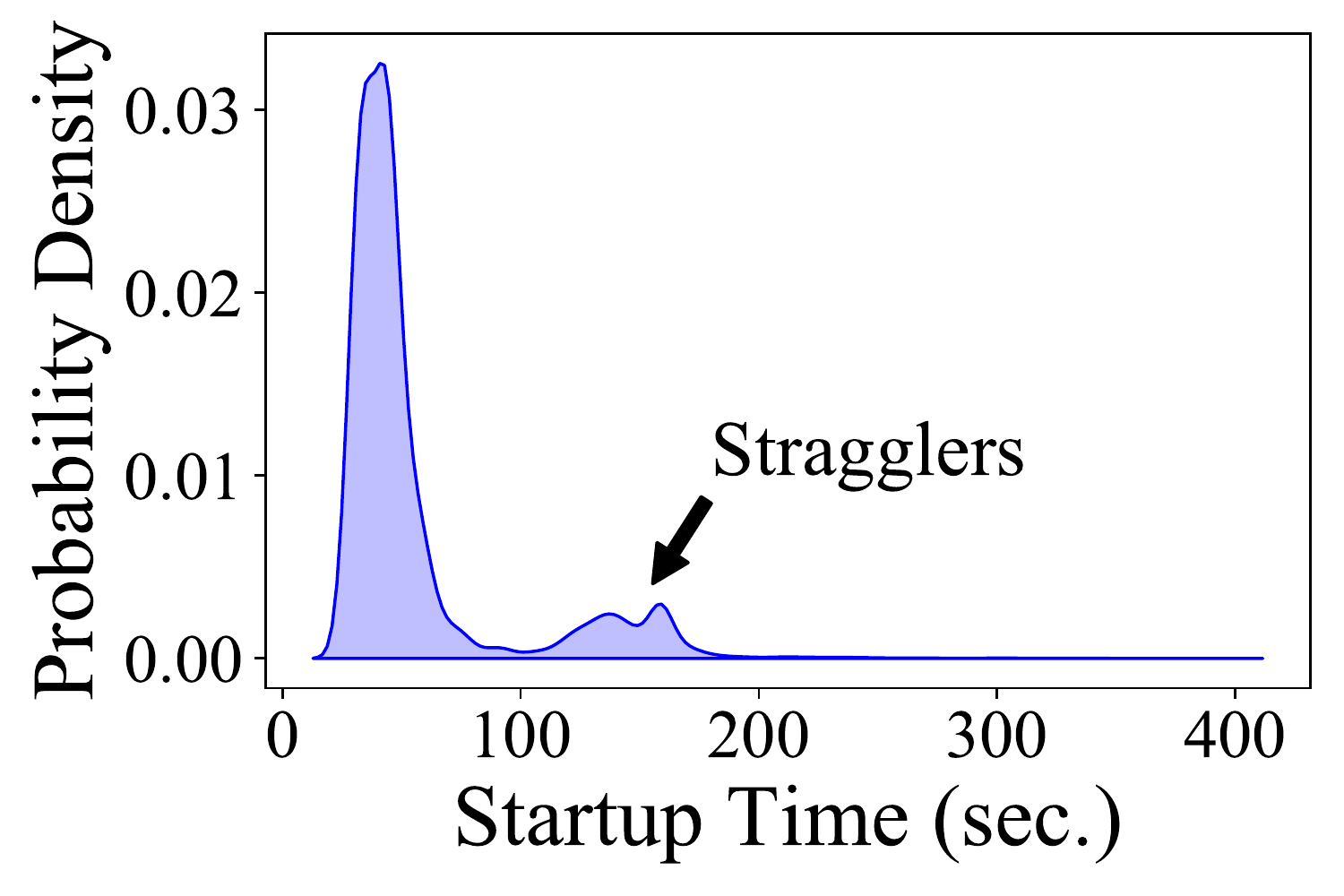}
	  }
	  \hfill
	  \subfloat[Warm Startup: Linux VM (256G)\label{fig:aws-vm-cold-w-256g}]{%
	    \includegraphics[width=0.47\columnwidth]{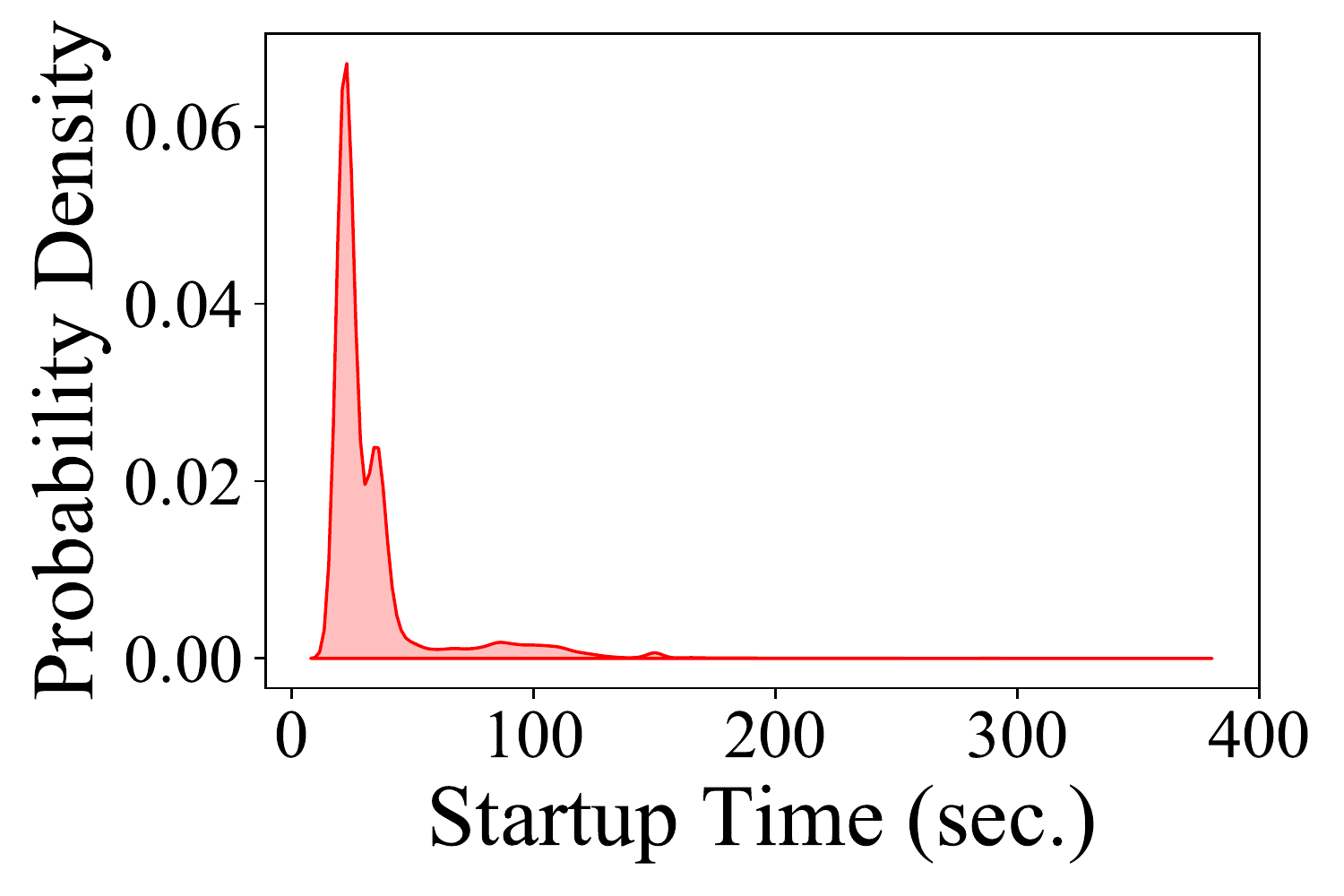}    
	  }
	  \caption{Startup time distribution of AWS Linux VMs}
	  \label{fig:aws-vm-dist}
    \end{minipage}
\end{figure}

Table~\ref{tab:all_startup} reports the average VM startup times of both Linux and Windows instances in AWS and GCP. For the cold startup times, {AWS VMs had shorter startup times than the VMs in GCP.} In particular, GCP VMs showed 2.22$\times$, (Linux VMs) and 1.2$\times$ (Windows VMs) slower startup times compared to the AWS VMs. Regarding the warm startup times, both providers showed similar startup times, and warm startup time is much shorter than cold startup time. The shorter time of the warm startup case is mainly because the VM images are already stored in the physical machine so that the warm startup process does not have any delays from VM image transfer.

Regarding OS impacts for VM startup times, we confirmed that {OS types are still impacting the VM startup times} based on the measurement result that both cloud providers showed different VM startup times between Linux and Windows VMs. For example, Linux VMs in AWS showed (40\%) faster startup times than Windows VMs, whereas, in GCP, Windows VMs had (13\%) faster startup times than Linux VMs.

\vspace{1mm}
\begin{mdframed}[backgroundcolor=blue!5] 
{\bf Observation\#1: Different OS types}

\noindent
$\bullet$ AWS VMs have faster startup time VMs in GCP.

\noindent
$\bullet$ In AWS, the startup times of Linux VMs are 40\% faster than Windows VM, but GCP shows the opposite results.

\end{mdframed}

\begin{figure}[t]
\centering
	\begin{minipage}[b]{1\columnwidth}
		\includegraphics[width=\columnwidth]{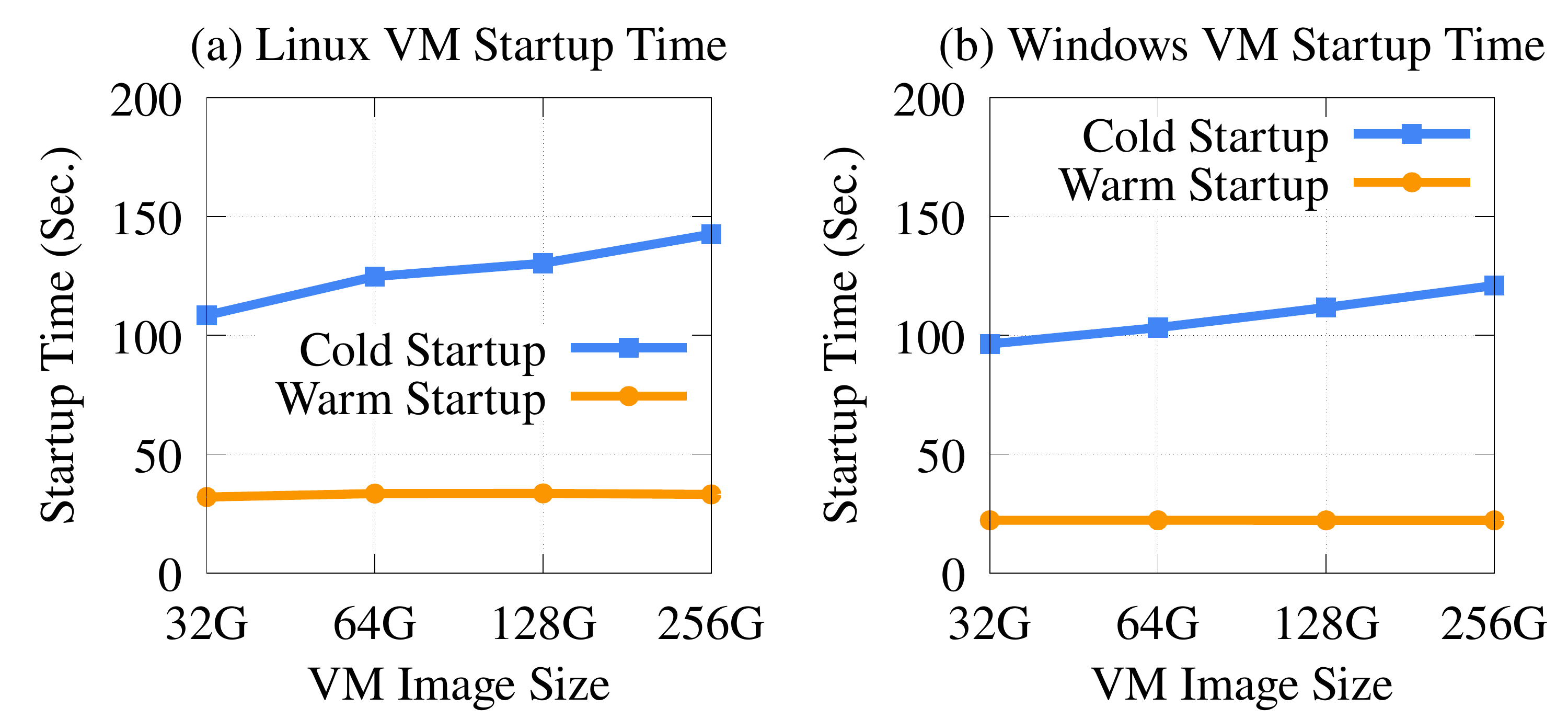}
		\caption{(GCP) VM startup times with different VM image sizes}
		\label{fig:gcp_vm_size}
    \end{minipage}

	\vspace{2.00mm}
  
  \begin{minipage}[b]{1\columnwidth}
  \subfloat[Cold Startup: Linux VM (128G)\label{fig:gcp-vm-cold-l-128g}]{%
    \includegraphics[width=0.47\columnwidth]{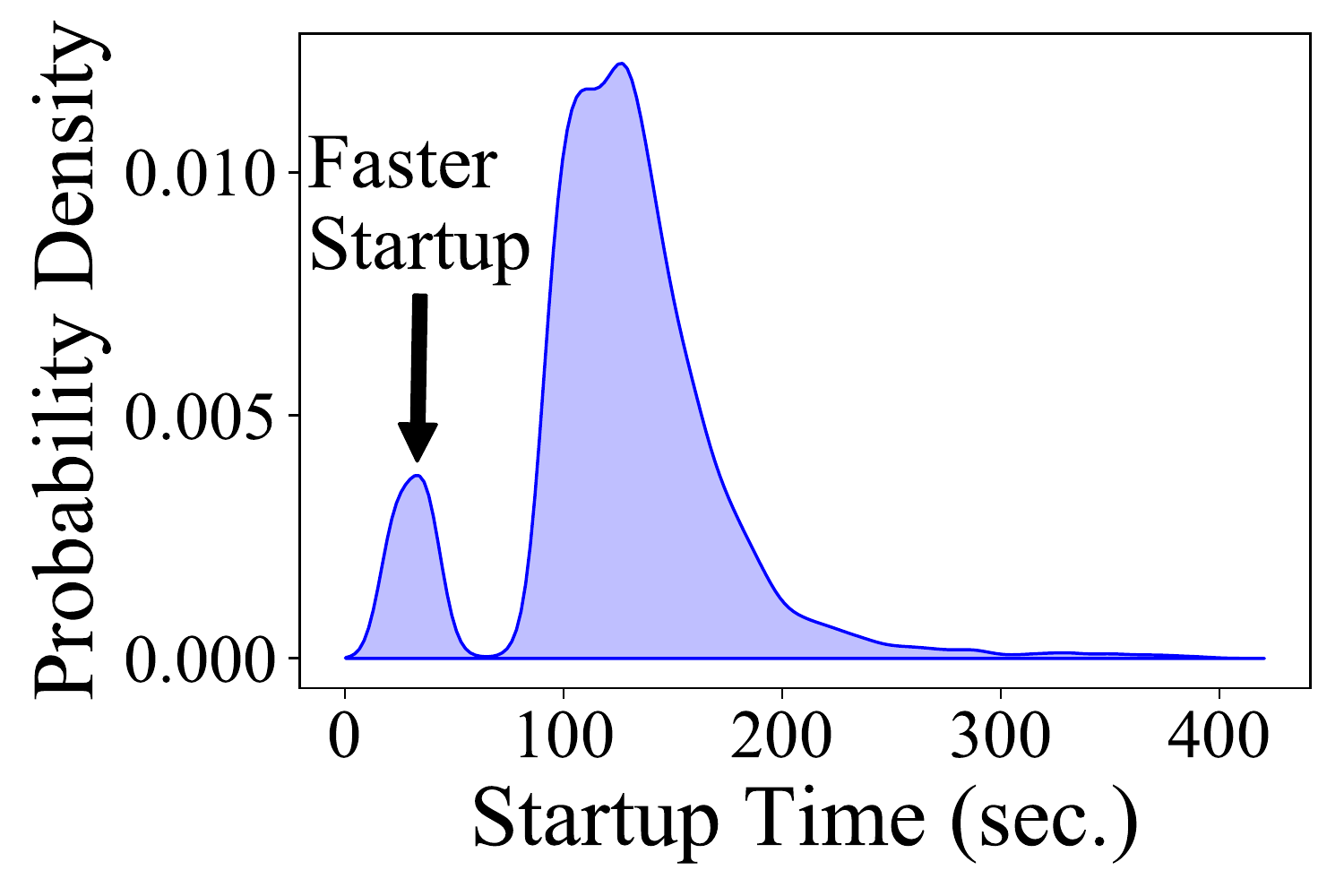}
  }
  \hfill
  \subfloat[Warm Startup: Linux VM (64G)\label{fig:gcp-vm-cold-w-64g}]{%
    \includegraphics[width=0.47\columnwidth]{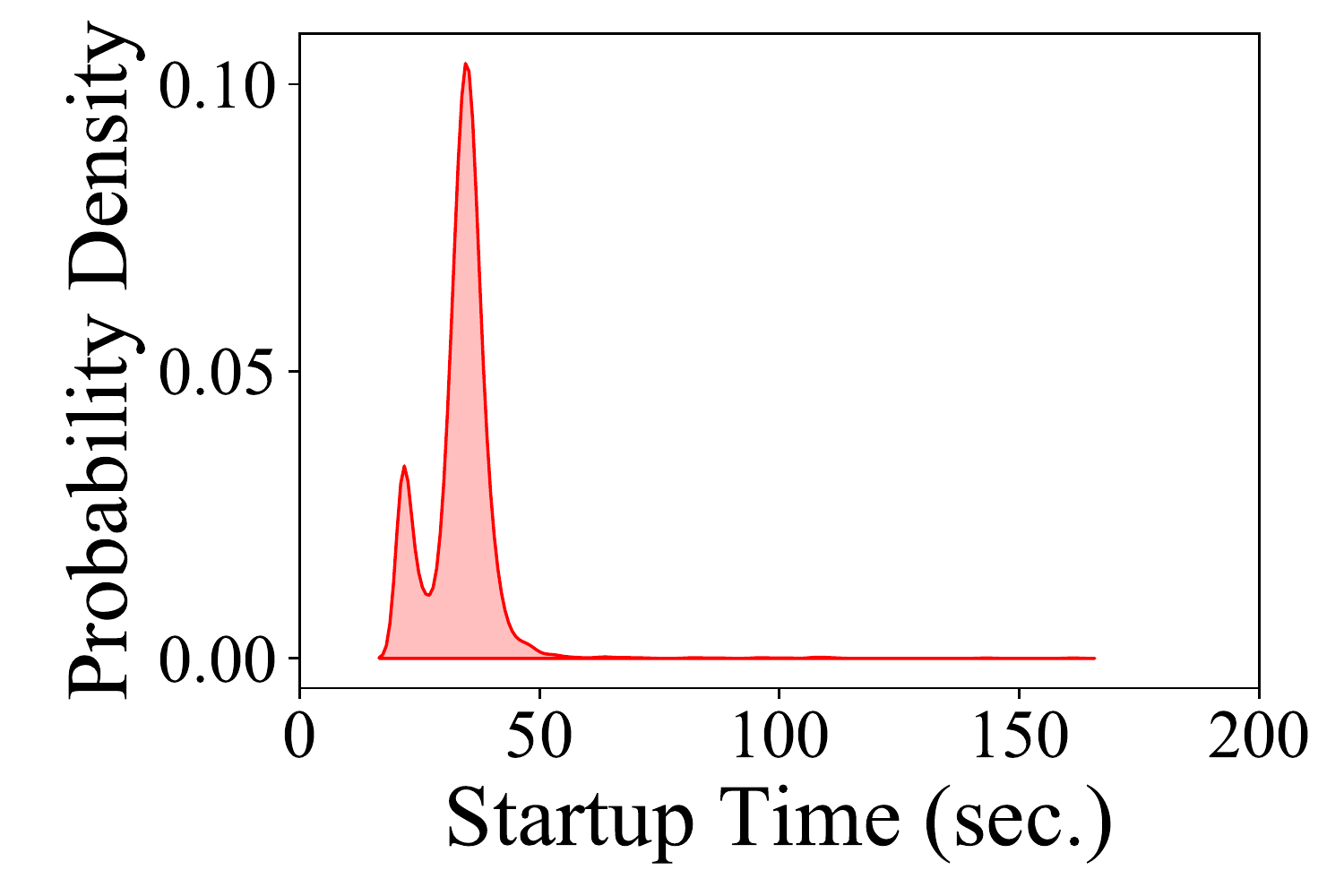}
  }
  \caption{Startup time distribution of GCP Linux VMs}
  \label{fig:gcp-vm-dist}
	\end{minipage}
\end{figure}

\subsection{VM Image Sizes}\label{subsec:img_size}

VM image size is another widely accepted factor regarding the variability in the VM startup times~\cite{StartupMao:CLOUD12}.
We measure VM startup times with various VM image sizes (from 32GB to 256GB) in AWS and GCP to confirm that VM image size is still changing startup times.

\FIG~\ref{fig:aws_vm_size} reports average startup times of AWS Linux and Windows VMs with four different VM image sizes. 
The results also show both cold and warm startup times of the VMs.
Unlike the previous findings, our measurement results confirm that 
{AWS VMs had almost constant startup time regardless of their image sizes.} 
We observed such constant patterns from both OS systems as well as both warm and cold startup times.
The maximum startup time difference between the smallest (32GB) and the largest (256GB) image sizes is only 2\% --  3\% (Linux VMs). {This result is clearly different from the widely used assumption that VM startup processes take longer as VM image size increases.} We assume that AWS could leverage several optimizations (described in Section~\ref{sec:relatedwork}) to provide constant startup time regarding this improvement.

However, the results only show the average startup times, so we also analyzed the distributions of VM startup times with the same image sizes. As shown in Fig.~\ref{fig:aws-vm-dist}, we observed that the VM startup times had multi-modal distributions, indicating that there were (straggler) VMs that have slow startup times. 
Specifically, such straggler VMs were clearly observed in AWS VM's cold startup times (e.g., Fig.~\ref{fig:aws-vm-cold-l-64g}).
We think such slow VM startup times were affected by other factors that we investigate later in this section. 

Fig.~\ref{fig:gcp_vm_size} reports VM startup times with different image sizes measured from GCP.
The GCP results showed that GCP VMs' startup times had a positive correlation with the VM image sizes, which is different from the AWS results. 
As shown in the figure, the GCP VMs' cold startup time increased as the VM image size increased, and both Linux and Windows VMs showed the same patterns in the change of VM startup time. 
The measurement results with cold startup time confirm that {the VM image sizes are still impacting the VM startup times in GCP.} Unlike the cold startup times, the warm startup times of GCP VMs were constant and stable regardless of the VM image sizes. 
However, stable warm startup times are expected because cloud providers can reuse existing VM images when warm startups are performed. 

Fig.~\ref{fig:gcp-vm-dist} shows the distributions of both cold and warm startup times from GCP Linux VMs. 
An interesting observation from the distributions is that the GCP measurement results had bi-modal distributions. 
In particular, a group of VMs had faster VM startup times than the majority of measured VMs (shown in Fig.~\ref{fig:gcp-vm-cold-l-128g}). 
Our further analysis revealed that the group of VMs having faster startup times was because of GCP's caching mechanism, which is described below.

\begin{table}[t]
\centering
\caption{VM image caching period in GCP.}
\begin{tabular}{|c|c|}
\hline
{\bf GCP Region} & {\bf Cache Period (Minutes)} \\ \hline
{\tt us-east4} (N. Virginia) & 75 -- 95 \\ \hline
{\tt us-central1} (Iowa) & 75 -- 100 \\ \hline
{\tt us-west1} (Oregon) & 70 -- 85 \\ \hline
{\tt europe-west1} (Belgium) & 70 -- 95 \\ \hline
{\tt asia-southeast1} (Singapore) & 75 -- 85 \\ \hline
\end{tabular}
\label{tab:cache_period}
\end{table}

\vspace{1mm}
\noindent
{\bf VM Image Cache Period in GCP.} 
As one of the major contributions of this work, we found and analyzed GCP's {\em VM image cache mechanism} that effectively reduces the startup time and possibly offers near-constant VM startup time (regardless of its image size). Based on our measurements, we found that the cache-based approach works similarly to several VM startup time reduction mechanisms proposed by the research community~\cite{ScalabeVMDeployment:SC13, VMCaching:Europar2012, CooporativeCaching:TCC2018}. 
The below procedure explains the VM image caching mechanism used in GCP.
\begin{enumerate}
	\item If a user creates a VM in a data center (a zone in a region) based on a specific VM image, which has not been used for a certain period of time ({\bf cache period}) in the data center, the VM image is transferred from an image repository in GCP to the data center. And the VM is created based on the transferred VM image, and then, the image is stored in the data center. In this step, the VM startup time follows the pattern reported in \FIG~\ref{fig:gcp_vm_size}.
	\item If a user creates another VM based on the VM image (used in step 1) in the same data center {\em within the cache period}, the VM is created using the VM image stored in the data center. In this case, the cold startup time (VM creation time) is much faster than step 1 because there is no delay from the VM image transmission.
	\item If the VM image is no longer used over the cache period, the VM image is removed from the data center.
\end{enumerate}

\vspace{1mm}
The VM image caching is only beneficial when a user creates the VM based on a previously used VM image in the same data center (zone) within the cache period. If the user creates a VM, based on the same image, in the {\em different} data center (zone) of the same region, then this mechanism does not work even within the cache period. So it is important to know the exact cache period of storing VM images in GCP.
Table~\ref{tab:cache_period} reports the VM image cache period in the five GCP regions. These results also include the cache period in all zones in the five regions. As shown in Table~\ref{tab:cache_period}, GCP data centers generally have 70 -- 100 minutes of VM image cache period. Within this period, the cold startup time of a VM with a pre-stored (cached) image is much faster than the startup time without cache.

\begin{figure}[t]
\centering
	\includegraphics[width=1\columnwidth]{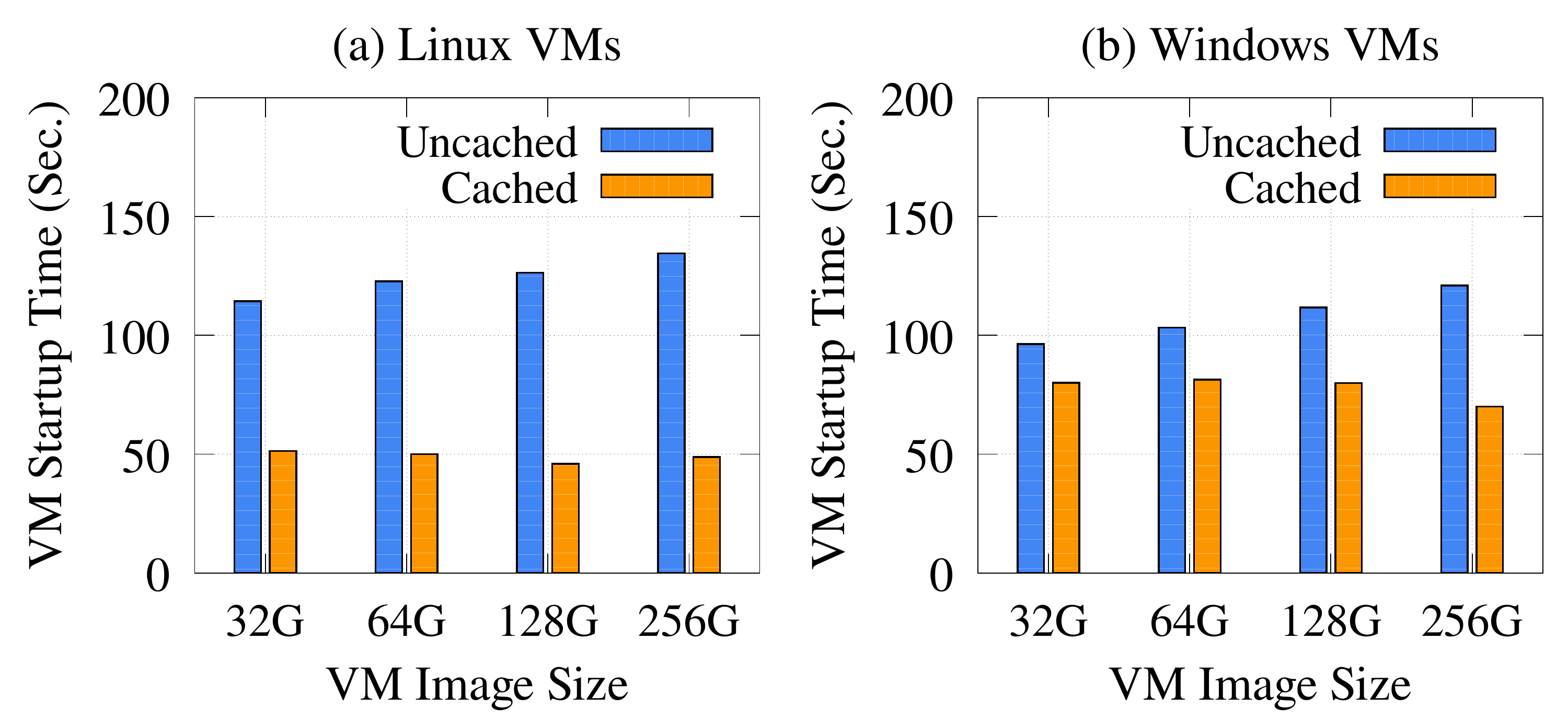}
	\caption{(GCP) Startup time differences with cached images.}
	\label{fig:google_cache}
\end{figure}

\FIG~\ref{fig:google_cache} shows the difference in VM cold startup time with and without cached VM image.
VMs with cached images showed much faster cold startup times compared to the uncached cases. With this cache mechanism, VM can reduce startup time by 60\% (Linux VM) and 27\% (Windows VM).

\vspace{1mm}
\begin{mdframed}[backgroundcolor=blue!5] 
{\bf Observation \#2: Different VM image sizes}

\noindent
$\bullet$ AWS VMs show near-constant VM startup times regardless of their image sizes. (However, there are some straggler VMs.)

\noindent
$\bullet$ In GCP, VM image sizes still impact the VM startup times. However, the startup times can decrease with GCP's internal caching mechanism. GCP's data centers keep recently created VM images for the next 75-10 minutes; thus, GCP can provide near-constant startup times when recreating VMs (based on the same image) within the cache period.

\end{mdframed}

\subsection{Instance Types}\label{subsec:instance_type}

\begin{figure}[t]
\centering
	\begin{minipage}[b]{1\columnwidth}
		\includegraphics[width=1\columnwidth]{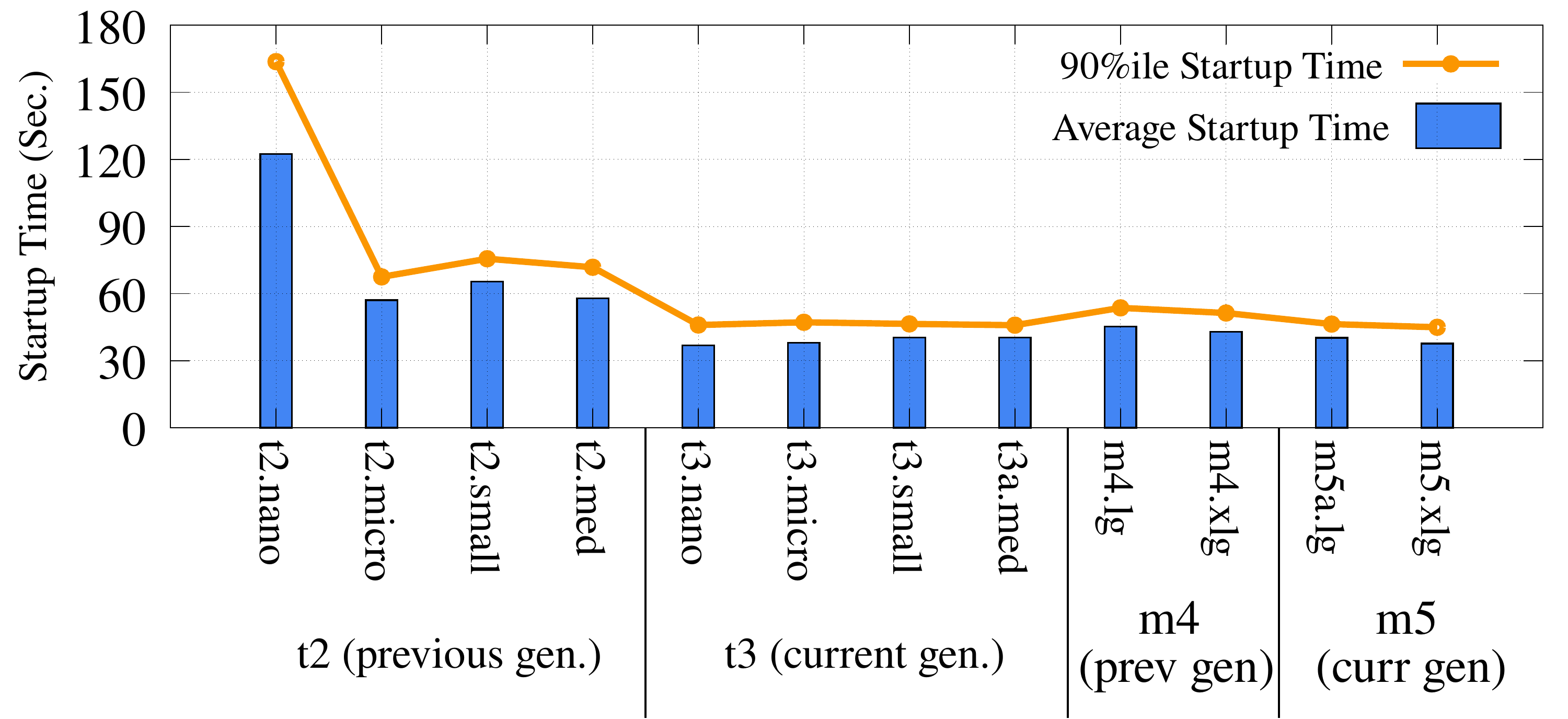}
		\caption{(AWS) Cold startup times of different VM types. This graph contains measurement results from on-demand Linux VMs (with all image sizes).}
		\label{fig:aws_vmtype}
	\end{minipage}

	\vspace{2.00mm}

	\begin{minipage}[b]{1\columnwidth}
		\includegraphics[width=1\columnwidth]{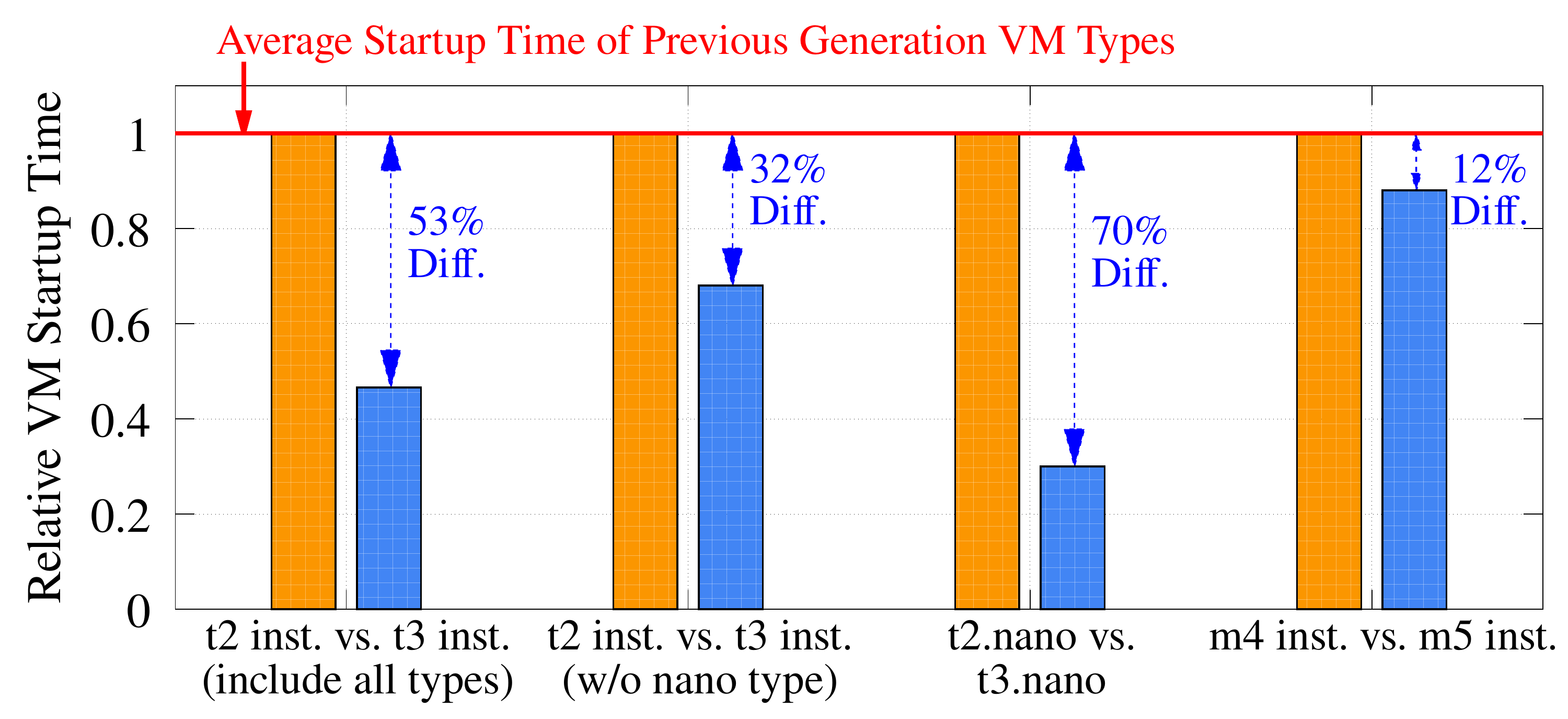}
		\caption{(AWS) Cold startup time comparison of VMs between older and newer VM types. This graph only contains measurement results from on-demand Linux VMs.}
		\label{fig:aws_vmtype_gen}
    \end{minipage}
\end{figure}

We also measured VM startup time variations of different VM types. As listed in Table~\ref{tab:vmtypes} (in Section~\ref{sec:methdology}), we used 12 VM types from AWS and 11 VM types from GCP. These instance types include shared core (or burstable) and general-purpose instances. \FIG~\ref{fig:aws_vmtype} reports the average and 90\%ile cold startup time variations of 12 instance types in AWS. Please note that the results are only reporting the cold startup times of different VM types with Linux OS, and all image sizes are considered for calculating the results as the VM image sizes do not meaningfully impact the startup times. 
We omit the results from Windows instances, and the results with Windows VMs had a similar pattern to the results reported in \FIG~\ref{fig:aws_vmtype}. 
{Among AWS instances, {\tt t2} instances (previous generation of burstable instances) showed significantly longer startup time compared to other instance types.} {\tt t2} instances had 85.15 seconds of average and 158.22 seconds of 90\%ile of startup time. In particular, {\tt t2.nano}, the smallest instance type in AWS (with 1 vCPU and 0.5GB memory), showed the most unstable and longest cold startup time among 12 AWS instances. For other instance types ({\tt t3}, {\tt m4}, and {\tt m5}), the VM types tended to have similar cold startup time with other VM types in the same instance family. 

We also observed that different generations in the same purpose instance types (e.g., {\tt t2} vs. {\tt t3}, {\tt m4} vs. {\tt m5}) could have differences in VM startup time. We measured the differences in cold startup times between older and newer generation instances. \FIG~\ref{fig:aws_vmtype_gen} shows the comparison results between {\tt t2} and {\tt t3}, as well as {\tt m4} and {\tt m5} instances. Regarding the comparison between {\tt t2} and {\tt t3}, we also report the results with including, excluding, and solely comparing nano types because {\tt t2.nano} showed substantially slower startup time. As shown in \FIG~\ref{fig:aws_vmtype_gen}, {the newer generation instances showed faster cold startup times than the older instance types in AWS}. For example, {\tt t3} instances had 53\% to 32\% (without {\tt nano} types) faster startup times compared to {\tt t2} instances. Specifically, the average startup time of {\tt t3.nano} was 70\% faster than that of {\tt t2.nano}, and {\tt t2.nano} only took 36.87 seconds on average. Regarding the general-purpose instances ({\tt m4} and {\tt m5}), the time differences between two instance families are smaller than {\tt t}-instances, and the results showed that {\tt m5} (newer-generation) instances had 12\% faster cold startup time than {\tt m4} (older-generation) instances.

\begin{figure}[t]
\centering
	\begin{minipage}[b]{1\columnwidth}
		\includegraphics[width=\columnwidth]{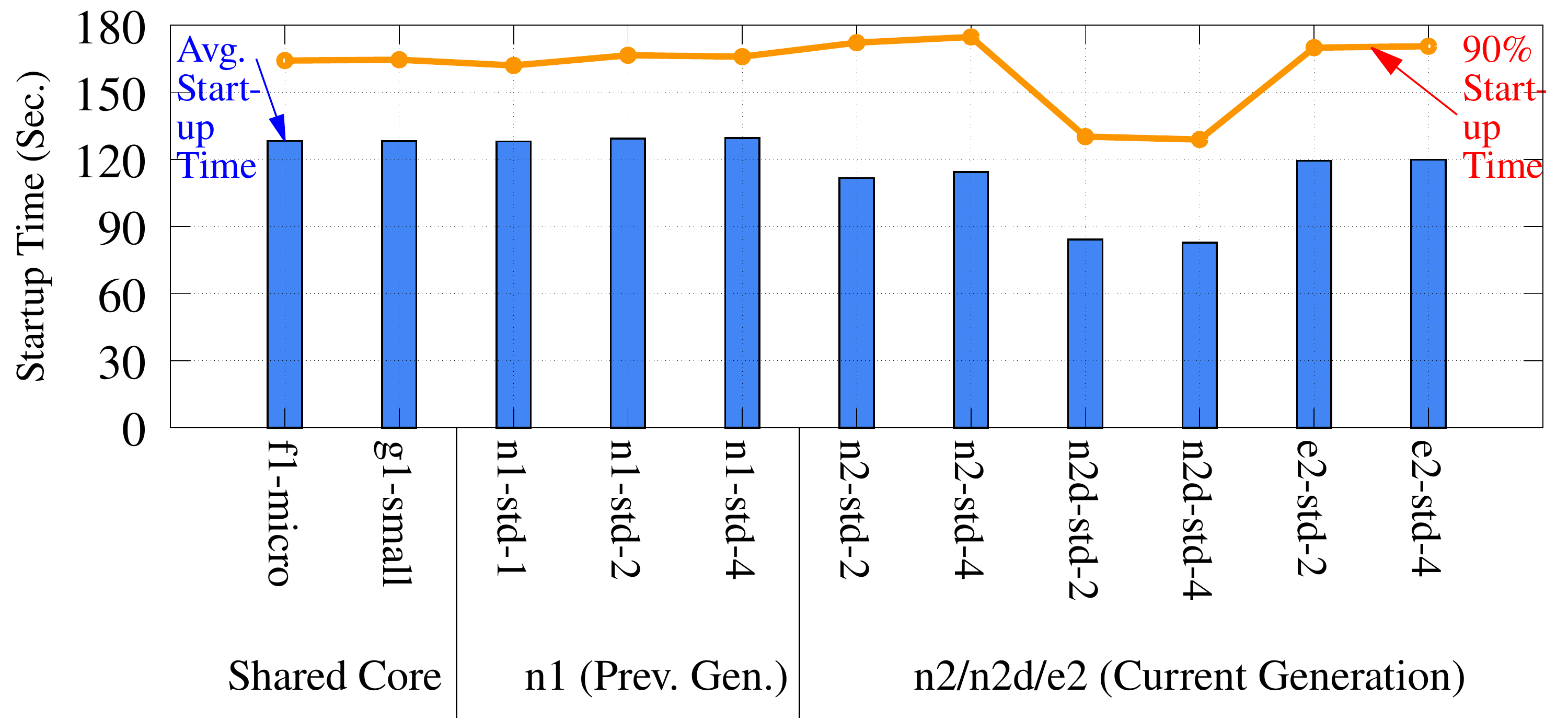}
		\caption{(GCP) Cold startup times of different VM types with Linux OS (with 64GB image size).}
		\label{fig:google_vmtype_linux}
	\end{minipage}

	\vspace{2.00mm}

	\begin{minipage}[b]{1\columnwidth}
		\includegraphics[width=\columnwidth]{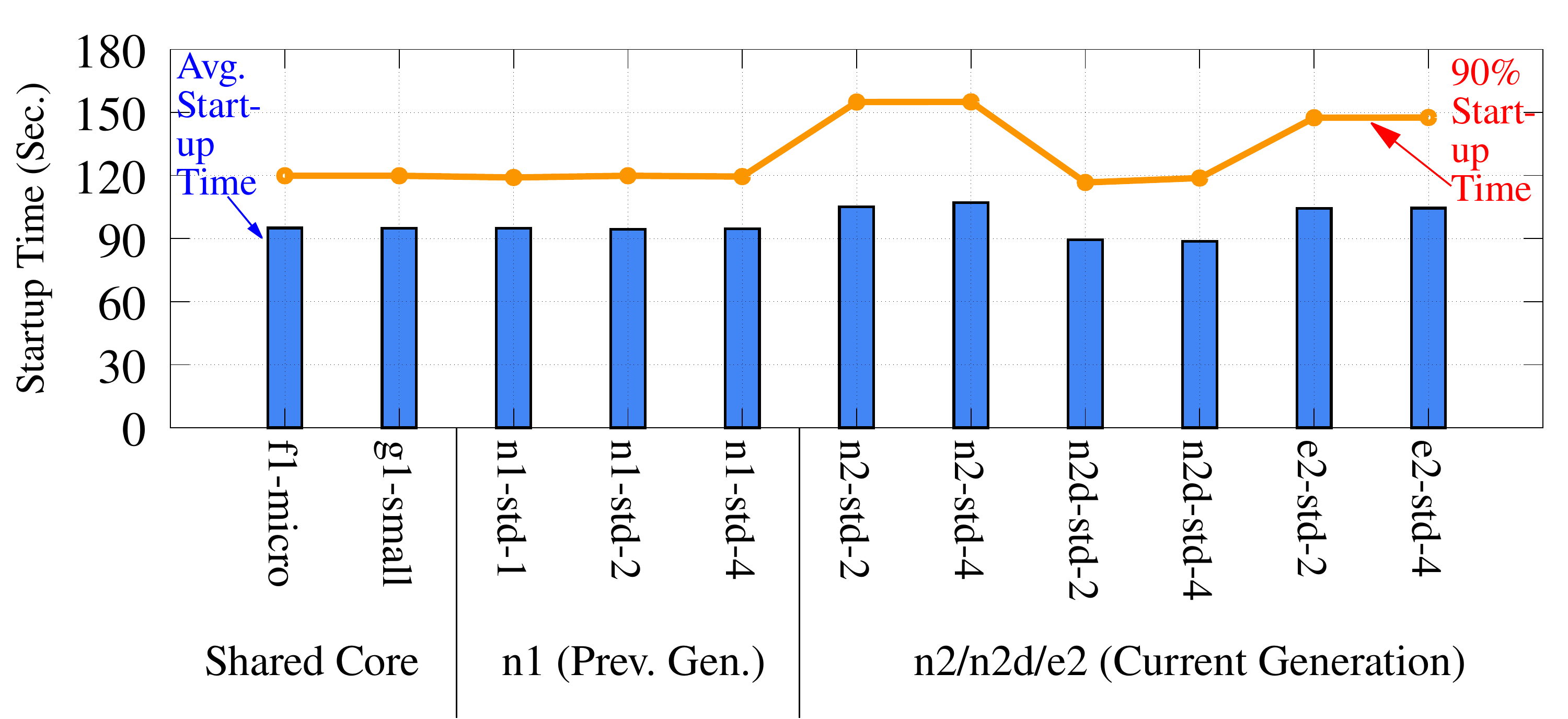}
		\caption{(GCP) Cold startup times of different VM types with Windows OS (with 32GB image size).}
		\label{fig:google_vmtype_windows}
    \end{minipage}
\end{figure}

Fig.~\ref{fig:google_vmtype_linux} and~\ref{fig:google_vmtype_windows} show the cold startup time variations of 11 VM types in GCP. Fig.~\ref{fig:google_vmtype_linux} contains the measurement results from Linux VMs with 64GB image size, and  Fig.~\ref{fig:google_vmtype_windows} has the measurement results from Windows VMs with 32GB image size. 
We omit the results with other image sizes because the measurements with other image sizes showed similar results with Fig.~\ref{fig:google_vmtype_linux} and~\ref{fig:google_vmtype_windows}. Regarding VM types with Linux OS (Fig.~\ref{fig:google_vmtype_linux}), share-core ({\tt f1-small} and {\tt g1-micro}) and {\tt n1\footnote{{\tt n1} instances are the first generation general-purpose machine type in GCP~\cite{GoogleCloud-MachineTypes}. {\tt n1} instances use one of the following Intel CPU models; Skylake, Broadwell, Haswell, Sandy Bridge, and Ivy Bridge CPU }} instances showed slower (up to 42\%) VM startup time than {\tt n2}, {\tt n2d}, and {\tt e2-standard} instances\footnote{{\tt n2}, {\tt n2d}, and {\tt e2-standard} instances are the second-generation general purpose VM types in GCP~\cite{GoogleCloud-MachineTypes}. {\tt n2} VMs run on Intel Cascade Lake CPUs, {\tt n2d} VMs are based on AMD EPYC Rome processors, and {\tt e2} instances are running on available CPU platforms from either Intel or AMD.}. In other words, {the first (previous) generation VM types ({\tt f1}/{\tt g1}/{\tt n1}) had longer startup time compared to the second (current) generation VM types ({\tt n2}/{\tt n2d}/{\tt e2}) with Linux OS in GCP}.

\begin{figure}[t]
	\centering
	\begin{minipage}[b]{1\columnwidth}
		\includegraphics[width=1\columnwidth]{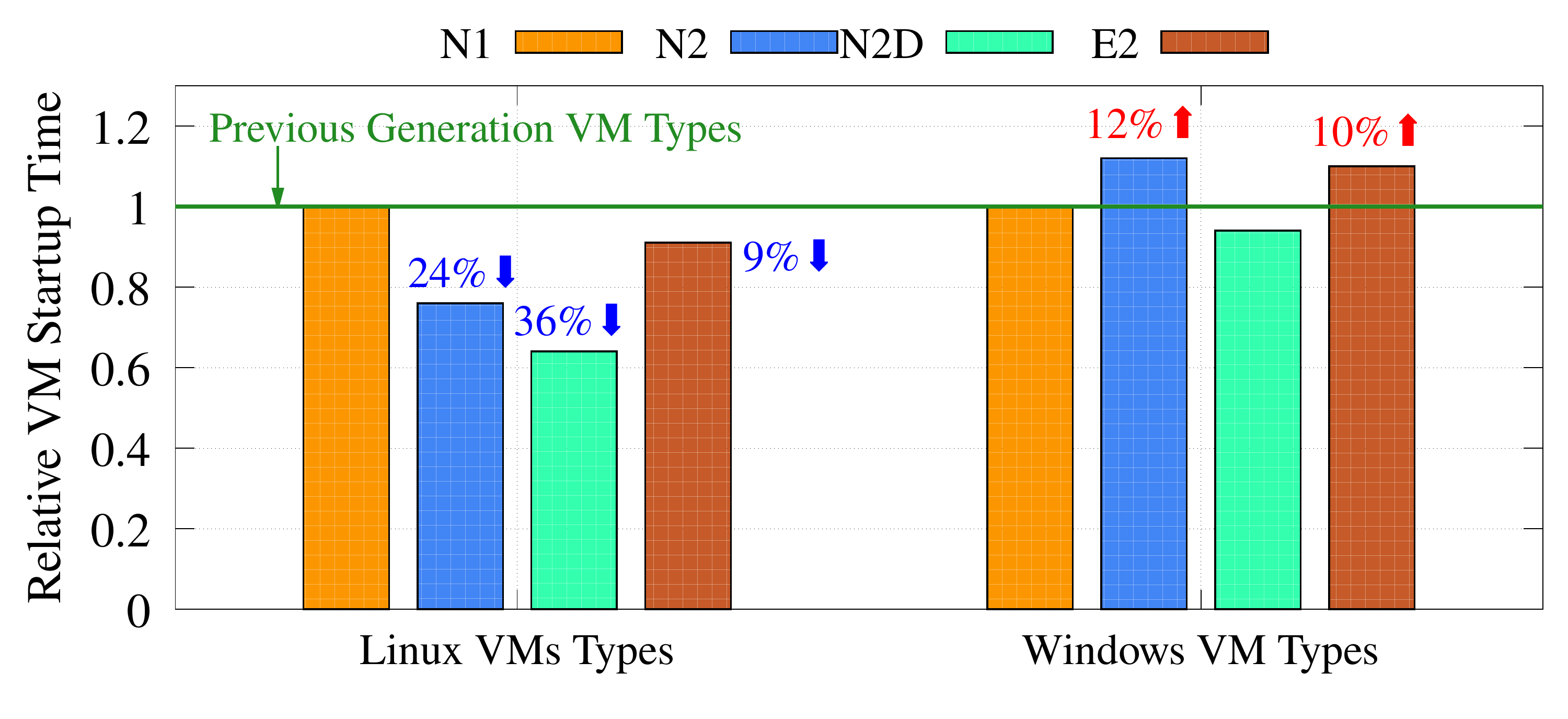}
		\caption{(GCP) Cold startup time comparison of (on-demand) VMs between older and newer VM types. }
		\label{fig:google_vmtype_gen}
	\end{minipage}

	\vspace{4.00mm}

	\begin{minipage}[b]{1\columnwidth}
		\includegraphics[width=1\columnwidth]{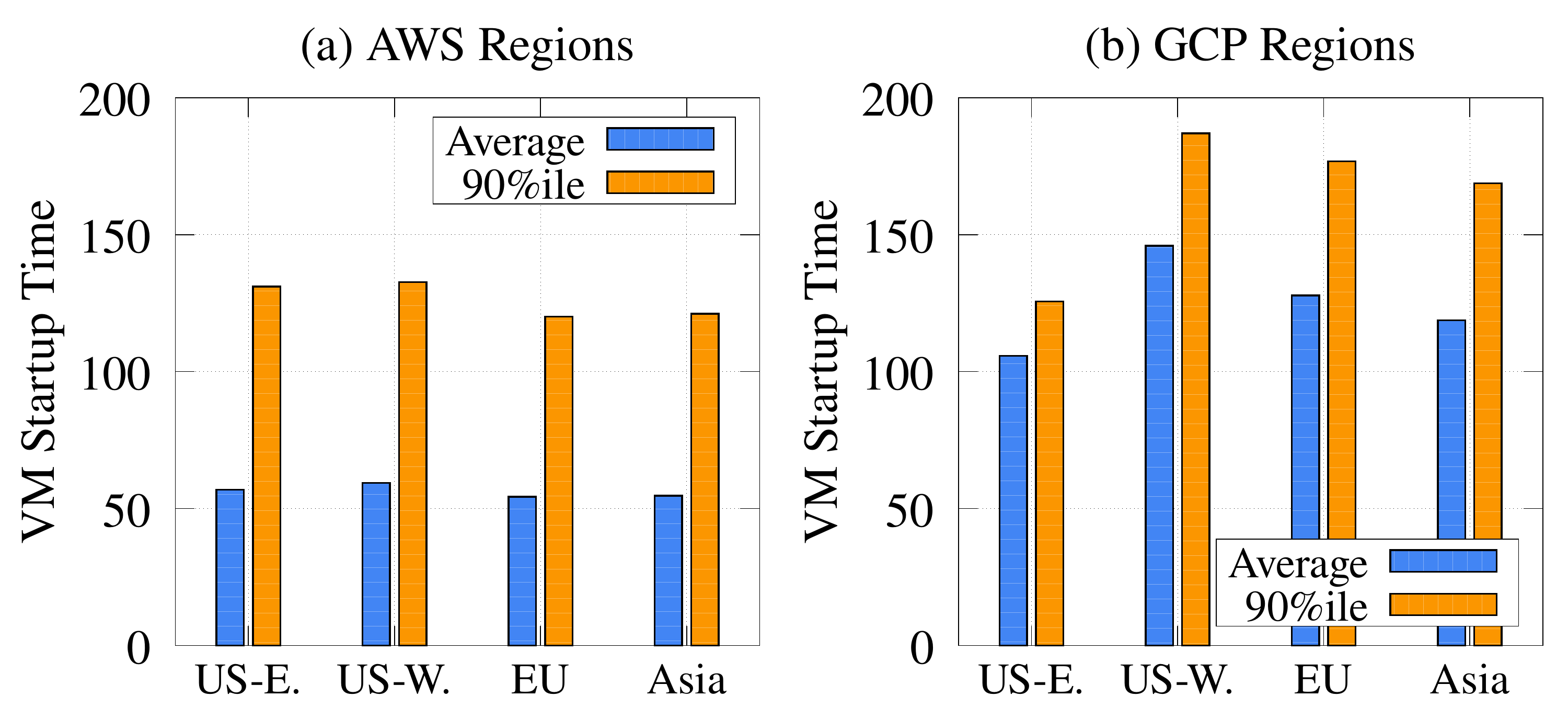}
		\caption{Average and 90\%ile VM cold startup times in different regions. (a) AWS VM startup times in four regions. US-E.: {\tt us-east-1} (N.Va), US-W.: {\tt us-west-2} (Oregon), EU: {\tt eu-west-3} (Paris), and Asia: {\tt ap-southeast-1} (Singapore). (b) GCP VM startup times in four regions. US-E.: {\tt us-east4} (N.Va), US-W.: {\tt us-west1} (Oregon), EU: {\tt europe-west1} (Belgium), and Asia: {\tt asia-southeast1} (Singapore).}
		\label{fig:byregion}
	\end{minipage}
\end{figure}

However, the measurement results from Windows VMs (Fig.~\ref{fig:google_vmtype_windows}) are different from the results with Linux VMs (Fig.~\ref{fig:google_vmtype_linux}). While the {\tt n2D} (second generation) instances showed the fastest VM startup times, {the first generation Windows VMs (e.g., shared-core and {\tt n1} instances) had shorter startup time than other second generation ({\tt n2}/{\tt e2}) instances}. The cold startup time differences between the first and second generation VMs in GCP are summarized in Fig.~\ref{fig:google_vmtype_gen}. In general, the newer generation instance types are 9\% -- 36\% faster in startup time than the older generation instance types with Linux, but, the startup times from the newer generation VM types with Windows can be slower (10\% -- 12\%) than the startup times from the older generation VM types.

\vspace{1mm}
\begin{mdframed}[backgroundcolor=blue!5] 
{\bf Observation \#3: Different instance types}

\noindent
$\bullet$ In AWS, the {\tt t2} family shows significantly longer startup time than other instance types. Specifically, {\tt t2-nano} is the slowest instance among 12 VM types.

\noindent
$\bullet$ In AWS, newer generation VM types (e.g., {\tt t3} and {\tt m5} family) show 12\% -- 70\% faster startup time than old generation VM instance types (e.g., {\tt t2} and {\tt m4}). 

\noindent
$\bullet$ In GCP, older generation ({\tt f1/g1/n1}) VMs with Linux OS show slower startup times than newer generation VMs (e.g., {\tt n2/e2}), but Windows VMs report the opposite results.
\end{mdframed}

\begin{figure}[t]
	\centering
	\begin{minipage}[b]{1\columnwidth}
		\includegraphics[width=1\columnwidth]{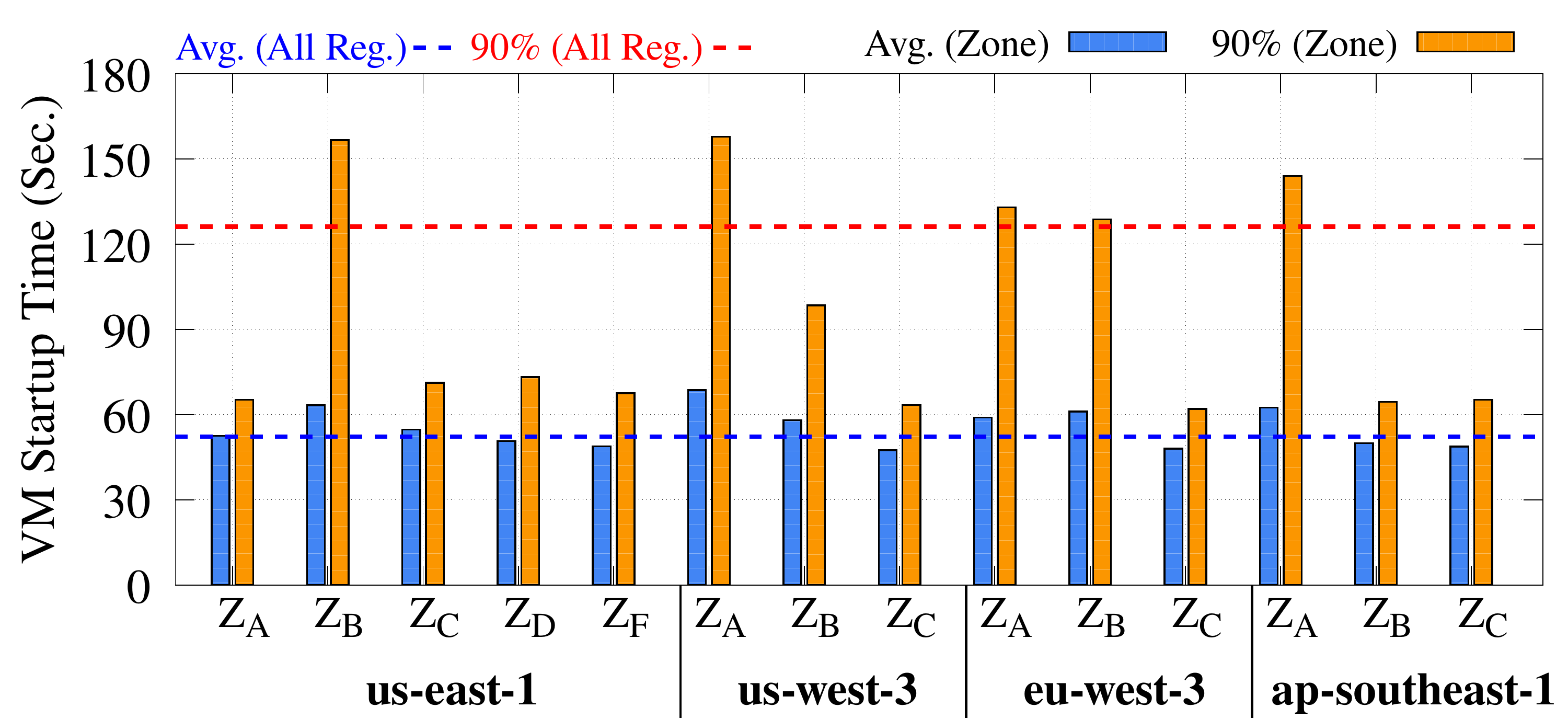}
		\caption{(AWS) Average and 90\%ile VM cold startup times in 14 different zones. (Z\textsubscript{A} -- Z\textsubscript{F}: Zone A to Zone F.)}
		\label{fig:byzoneAWS}
	\end{minipage}

	\vspace{4.00mm}
	\begin{minipage}[b]{1\columnwidth}
		\includegraphics[width=1\columnwidth]{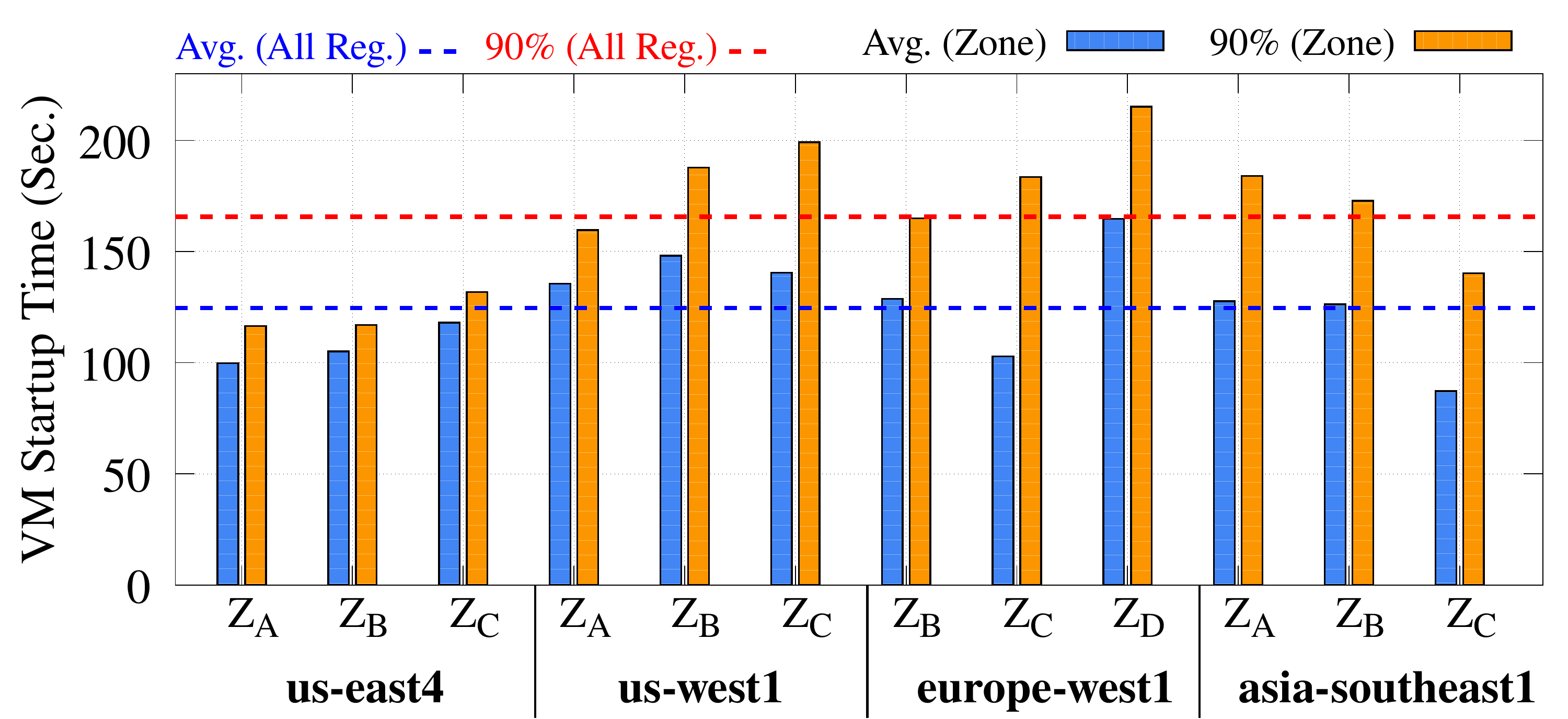}
		\caption{(GCP) Average and 90\%ile VM cold startup times in 12 different zones. (Z\textsubscript{A} -- Z\textsubscript{D}: Zone A to Zone D.)}
		\label{fig:byzoneGoogle}
	\end{minipage}
\end{figure}

\subsection{Regions and Data Center Locations}\label{subsec:data_loc}

In this subsection, we analyzed the VM startup time variations in different regions. \FIG~\ref{fig:byregion} shows the average and 90\%ile (cold) VM startup times in four different regions (U.S. east, U.S. west, EU, and Asia) from both providers. Please note that the results in \FIG~\ref{fig:byregion} were collected from on-demand Linux VMs. We omit the results from Windows VMs because they showed similar results with the Linux VMs. As shown in \FIG~\ref{fig:byregion}(a), the four regions in AWS had relatively consistent and stable VM cold startup times. The maximum difference in the four regions was only 5.02 seconds on average. However, the VM (cold) startup times (both average and 90\%ile) of GCP varied considerably with different regions. The maximum startup time difference was about 40 seconds on average, which is 32\% of average startup times of all four regions. Among four regions {\tt us-east4} (US-E., N.Va) was the fastest region and {\tt us-west1} (US-W., Oregon) showed the longest VM startup times.

We further analyzed the VM startup times per zones,
and \FIG~\ref{fig:byzoneAWS} reports the Linux VM startup time differences in 14 zones that belong to four AWS regions.
{Although AWS showed similar VM startup times in four regions (in our previous analysis), the VM startup times were fluctuating as per we choose different zones even within the same region}. Among 14 zones we measured in AWS, 3 zones (Z\textsubscript{A} in {\tt us-west-2}, Z\textsubscript{B} in {\tt us-east-1}, and Z\textsubscript{A} in {\tt ap-east-1}) had at least 20\% longer startup time compared to the average startup time of all four regions. VMs in another 3 zones (Z\textsubscript{B} in {\tt us-west-2}, Z\textsubscript{A} and Z\textsubscript{B} in {\tt eu-west-3}) had 11\% to 17\% longer startup time.
Moreover, it is observed that there was up to a 45\% difference (average startup time) among different zones, even within the same region ({\tt us-west-2}).
These results thus imply that to minimize VM startup time, AWS users should carefully choose zones when creating VMs.

\FIG~\ref{fig:byzoneGoogle} shows the VM startup time variations of Linux VMs in 12 different zones within four regions in GCP. The overall startup time variations are similar to the startup time fluctuations reported in \FIG~\ref{fig:byregion}(b). And, we also observed {there are significant startup time variations from a zone to another belonging to the same geographical region in GCP}.
For example, Z\textsubscript{d} in {\tt europe-west1} had 32\% longer VM startup time than the GCP average. Interestingly, the Z\textsubscript{c} of the same region had 3\textsuperscript{rd} fastest VM startup time (18\% shorter) out of all 12 zones, and the VM startup time difference in the {\tt europe-west1} region can be up to 62 seconds.
In particular, {\tt europe-west1} and {\tt asia-southeast1} showed high fluctuations of VM startup time among their zones. We observed that {\tt europe-west1} and {\tt asia-southeast1} regions could have 33\% -- 50\% differences in the VM startup times.
Consequently, these results also imply that GCP users need to carefully choose zones in a specific region when starting VMs for minimizing the VM startup time in order to improve resource elasticity.

\vspace{1mm}
\begin{mdframed}[backgroundcolor=blue!5] 
{\bf Observation \#4: Different data center regions}

\noindent
$\bullet$ In both cloud providers, different zones in the same region can significantly differ in VMs' startup time.
\end{mdframed}

\begin{figure}[t]
\centering
	\includegraphics[width=1\columnwidth]{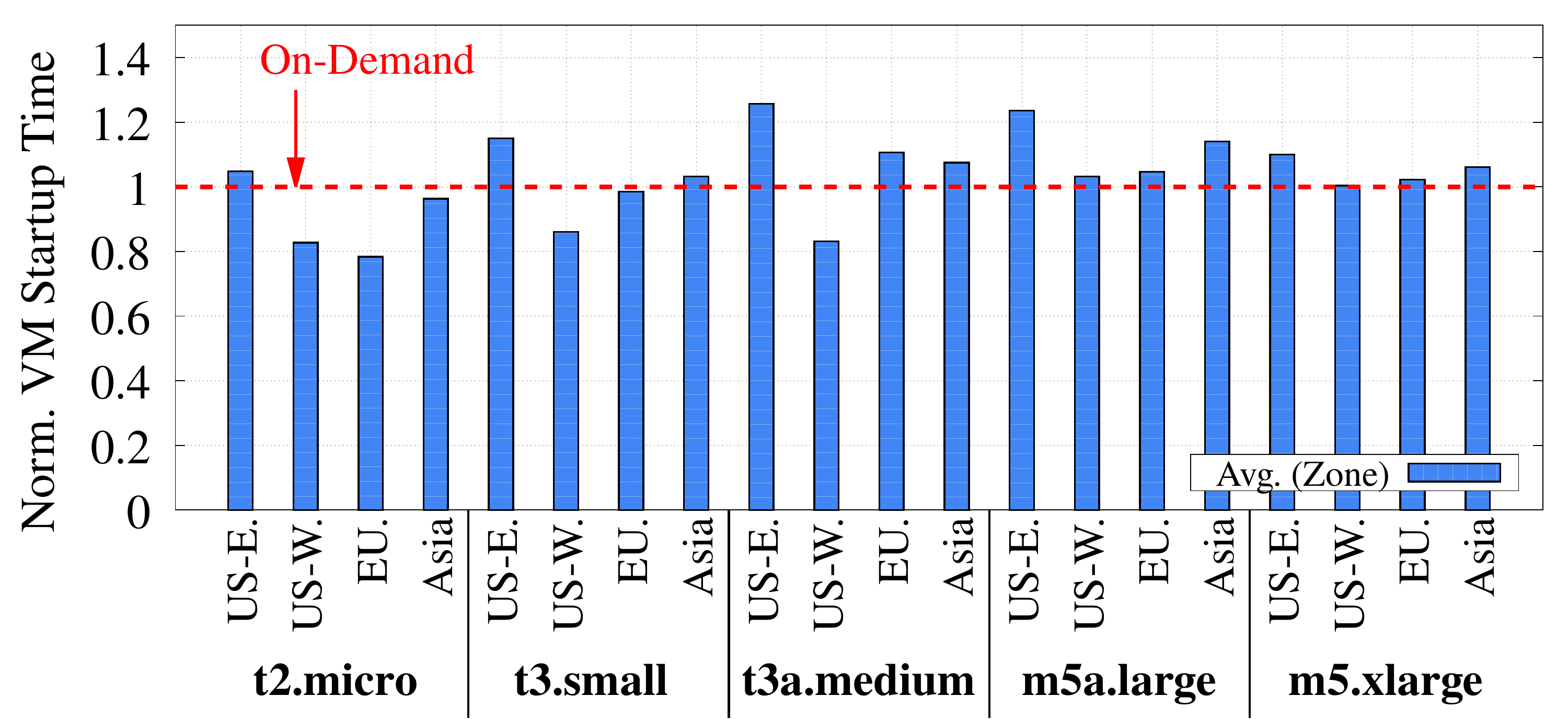}
	\caption{(AWS) Normalized startup times of spot instances. The startup times of spot VMs are normalized to the startup times of on-demand VMs.}
	\label{fig:ODSpotAWS}
\end{figure}

\subsection{Other Potential Factors}

In addition to previous analyses, we further investigated potential factors that can change the VM startup time. In this subsection, we provide our analysis of the VM purchase models and temporal factors for changing VM startup times.

\vspace{1mm}
\noindent
{\bf VM Purchase Models.} 
Public cloud providers typically offer three different ways of purchasing VMs, which are on-demand, reserved, and low-availability models. In particular, low-availability models, such as spot instances in AWS~\cite{AWS-Spot-online} and preemptible VMs in GCP~\cite{GoogleCloud-PVM-online}, are being increasingly used by cloud users due to the high cost-efficiency. As per a previous study~\cite{StartupMao:CLOUD12} and industry report~\cite{ZillowSpot:online}, the AWS spot instances tended to have longer cold startup times, so it is important to see if the previous reports are still valid. A potential factor for AWS spot instances having longer startup time is the bidding process to determine the VMs' price. 
However, in this study, we always used the bidding prices higher than the current price of spot instances to minimize the delay due to the bidding process and precisely measure the startup time or delay only caused by the cloud infrastructure. 
Moreover, in this measurement, we experienced the failure of VM startup (or provisioning) in both AWS spot and GCP preemptible VMs due to the limited capacity of such instances. Nevertheless, we do not report the failure rate or statistics of the startup process of the low-availability VMs as it is out of the scope of this work.

\begin{table}[t]
\centering
\caption{VM startup time differences between on-demand and preemptible VMs in GCP}
\begin{tabular}{|l|c|c|c|}
\hline
\textbf{Region} & \textbf{Zone} & \textbf{Linux VMs} & \textbf{Windows VMs} \\ \hline
\multirow{3}{*}{\tt us-east-4} & a & 0.6\% & 0.6\% \\ \cline{2-4}
 & b & 1.2\% & 0.9\% \\ \cline{2-4}
 & c & 0.9\% & 0.5\% \\ \hline
\multirow{3}{*}{\tt us-west-1} & a & 1.1\% & 2.3\% \\ \cline{2-4}
 & b & 1.9\% & 2.0\% \\ \cline{2-4}
 & c & 2.6\% & 5.2\% \\ \hline
\multirow{3}{*}{\tt europe-west1} & b & 5.2\% & 5.9\% \\ \cline{2-4}
 & c & 2.5\% & 5.5\% \\ \cline{2-4}
 & d & 0.9\% & 3.3\% \\ \hline
\multirow{3}{*}{\tt asia-southeast1} & a & 4.9\% & 4.7\% \\ \cline{2-4}
 & b & 3.2\% & 1.6\% \\ \cline{2-4}
 & c & 3.7\% & 5.9\% \\ \hline
\end{tabular}
\label{tab:google_preemptive}
\end{table}

\FIG~\ref{fig:ODSpotAWS} shows the startup times of the five different spot VM types, and the startup times are normalized to the startup time of on-demand VMs. While the startup time of spot instances (e.g., {\tt t2-micro} type) could be considerably (e.g., 20\%) different from the on-demand VMs' startup time, we could not observe any measurement results, which support that spot VMs had longer startup times than on-demand VMs. Interestingly, it is observed that spot VMs can have a shorter startup time than on-demand VMs. i.e., {\tt t2}/{\tt t3} instances in {\tt us-west-1}. Similar results (as reported in Table~4) were measured from GCP's preemptible VMs. The maximum difference in the startup time between preemptible and on-demand VMs was less than 6\% (Windows VMs in {\tt asia-southeast1}), and most zones/regions have less than 2\% -- 3\% difference in the startup times between two models. Also, we have observed several cases that the preemptible VMs had shorter (cold) startup time. As a result, {low-availability models (spot and preemptible VMs) no longer have slower startup time compared to on-demand VMs.} 

\vspace{1mm}
\noindent
{\bf Temporal Factors.} 
Cloud data centers may have a different amount of workloads as per ``time-of-day'' or ``day-of-week'' because cloud applications (e.g., web) may have repeating or diurnal workload patterns~\cite{PredEval:Cloud2016, CloudInsight:Cloud2018}. So, we further analyzed whether such temporal factors could change the VM startup times. We first examined the VM startup time variations on different days of week. While we omit the visualized results for this variation, both providers commonly showed that the VMs have shorter startup times on Saturdays and Sundays and longer startup times on weekdays. For example, AWS has a 14\% longer VM startup times on weekdays.

\begin{figure}[t]
\centering
		\includegraphics[width=\columnwidth]{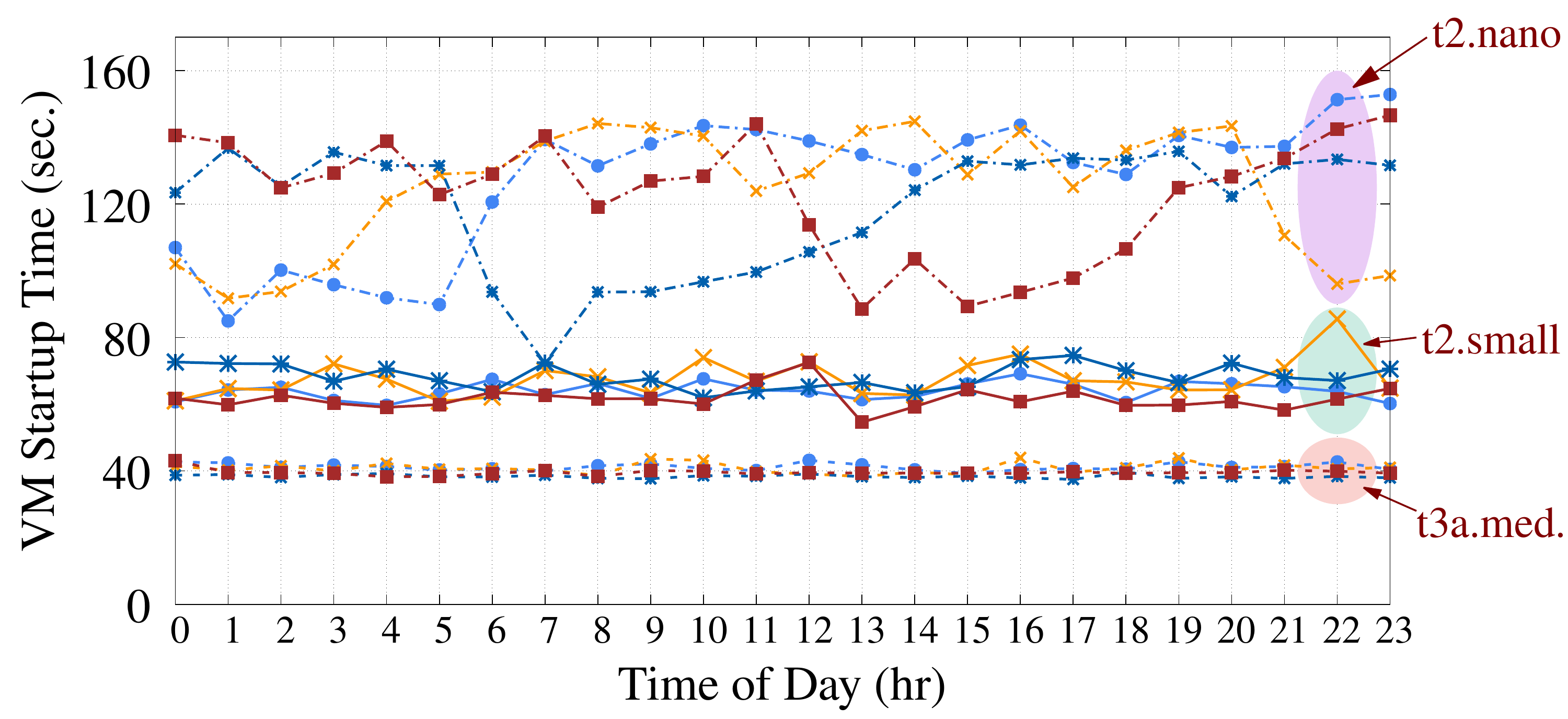}
		\caption{(AWS) VM startup time changes of three VM types ({\tt t2.nano}, {\tt t2.small}, and {\tt t3a.medium}) at different times of day. These results only contain the measurement results with Linux VMs.}
	\label{fig:byToDAWS}
\end{figure}

\begin{figure}[t]
\centering
        \includegraphics[width=\columnwidth]{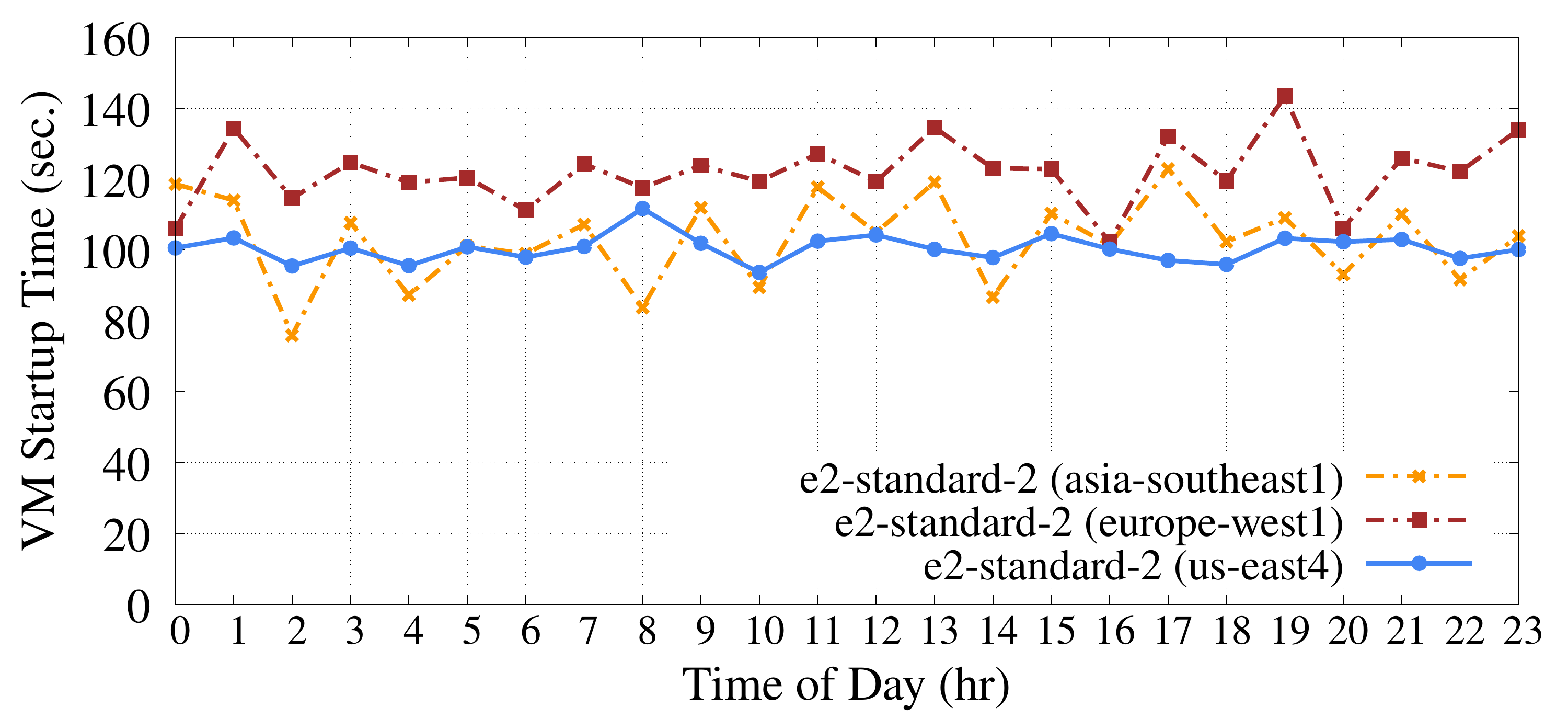}
        \caption{(GCP) VM startup time fluctuations of VMs ({\tt e2-standard-2} instance type) at different times of day. The results were measured with Linux VM with 32GB.}
    \label{fig:byToDGoogle}
\end{figure}

\FIG~\ref{fig:byToDAWS} reports the VM startup time differences of three VM types at a different time-of-day, and the results confirm that {the VM startup times could be varying over time}. We also observed that smaller and older generation VM types could have more substantial VM startup time changes compared to bigger and newer generations VMs. As shown in \FIG~\ref{fig:byToDAWS}, {\tt t2} instances (specifically {\tt t2.nano}) in AWS showed more dynamic changes in their startup times. \FIG~\ref{fig:byToDGoogle} shows the VM startup time changes in a GCP VM type in three different regions at different time-of-day. Similar to the AWS result, the startup times of GCP VMs were varying with time-of-day, but the startup times are more affected by location factors (e.g., geographical regions). As shown in \FIG~\ref{fig:byToDGoogle}, {\tt e2-standard-2} VMs in {\tt asia-southeast1} and {\tt europe-west1} had more significant changes in their startup times than VMs in {\tt us-east4}. The results reported in \FIG~\ref{fig:byToDAWS} and~\ref{fig:byToDGoogle} are consistent with our findings in Section~\ref{subsec:instance_type} and~\ref{subsec:data_loc}. Also, the results indicate that the VM startup times in both providers can be affected by temporal factors, and {the VM startup time fluctuations can be amplified by other factors, such as different VM types (in AWS) and geographical locations/regions (in GCP)}.

\vspace{1mm}
\begin{mdframed}[backgroundcolor=blue!5] 
{\bf Observation \#5: Other Potential Factors}

\noindent
$\bullet$ Spot or preemptible VMs (low availability models) no longer have slower startup time than on-demand VMs.

\noindent
$\bullet$ The temporal factors (different times) can change VMs' startup time. Moreover, VMs' startup time can be amplified by other factors like VM types and locations/regions.
\end{mdframed}

\section{Discussion}\label{sec:discussion}
\subsection{Comparison with Previous Measurement Results}

\begin{figure}[t]
	\centering
		\includegraphics[width=1\columnwidth]{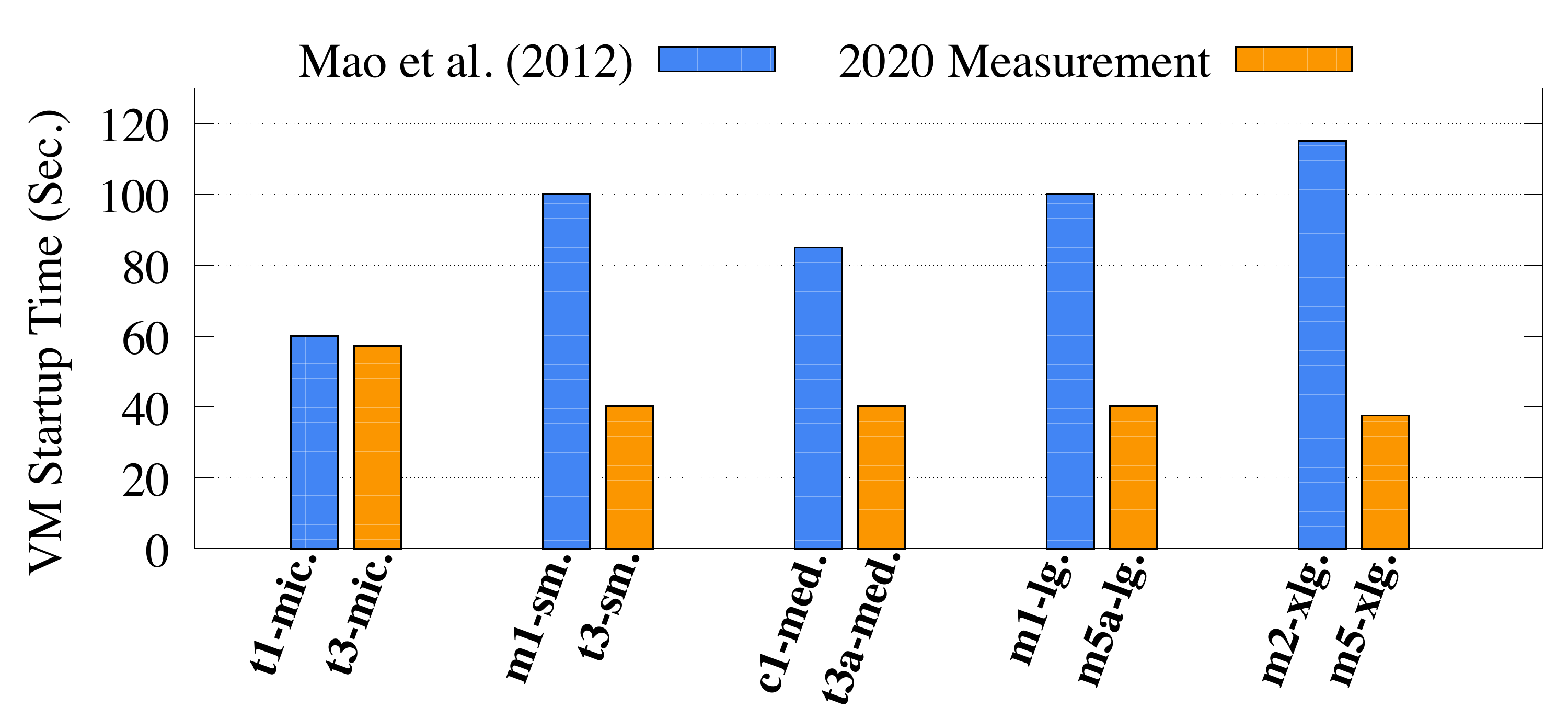}
		\caption{Comparison with previous work (Mao et al.~\cite{StartupMao:CLOUD12}); VM startup times with different VM types. Note that the VM image size in 2012 measurement was 0.5G and the results from this work contain all VM image sizes.}
		\label{fig:diss_vm_type}
\end{figure}

We compare our analysis results with previous reports to see how much improvement has been made by cloud providers. For the comparison, we use AWS's measurement results reported by Mao et al.~\cite{StartupMao:CLOUD12} in 2012 and measurements from both AWS and GCP by Abrita et al.~\cite{Provisioning:Bench18} in 2018. Please note that the results and graphs for Mao et al.'s measurements in this section are our interpretations of graphs and tables in the original author's report~\cite{StartupMao:CLOUD12}. Therefore, the results in \FIG~\ref{fig:diss_img_size} and~\ref{fig:diss_vm_type}, and Table~\ref{tab:google_preemptive} may have a marginal difference from the original results.

\subsubsection{VM Startup Time Comparison with Different VM Image Sizes} 
\begin{figure}[t]
	\centering
		\includegraphics[width=1\columnwidth]{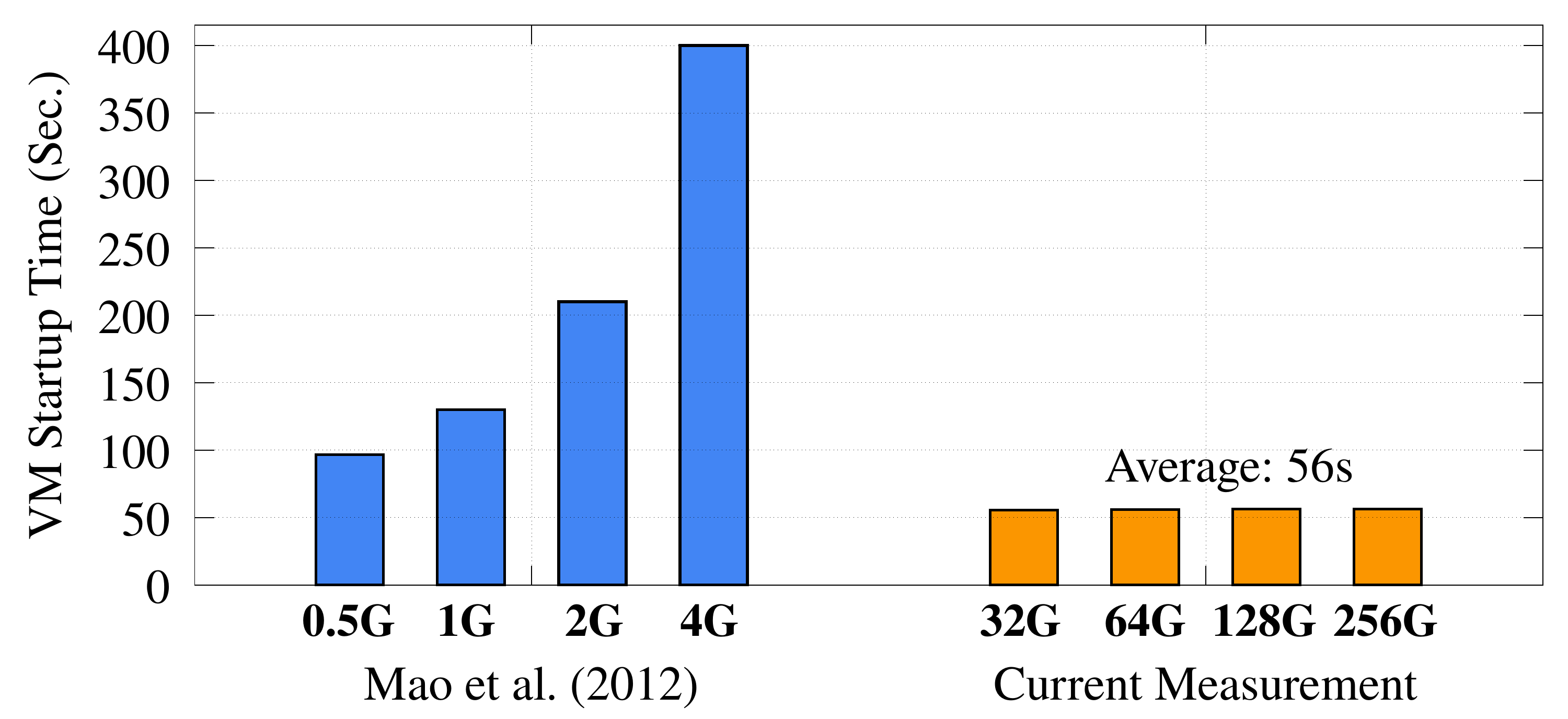}
		\caption{Comparison with previous work (Mao et al. in 2012~\cite{StartupMao:CLOUD12}); VM startup times with different VM image sizes.}
		\label{fig:diss_img_size}
\end{figure}

We first compare the VM startup times with various VM image sizes. 
\FIG~\ref{fig:diss_img_size} shows AWS's VM startup times measured by Mao et al. (2012) and by this work with different VM sizes. In the 2012 measurement, AWS clearly showed that the VM startup time has a positive correlation with the VM image sizes. Based on this result, we assume that AWS used external storage to store VM images, and inter/intra-data center networks may increase VM startup time. However, AWS in 2020 does not appear to have such a pattern in their VMs' startup time. AWS now can offer near-constant VM startup time with various image sizes. Moreover, the image sizes we used in this work are much larger than the VM image sizes used in 2012. e.g., up to 4G in 2012 vs. up to 256G in 2020. The measured startup time is much smaller than the previous measurement, implying that AWS significantly improved the data center infrastructures so that and AWS could dramatically reduce VM startup times. 

\subsubsection{VM Startup Time Comparison with Different VM Types}\label{subsubsec:compare_vmtype}
Next, we compare the VM startup times with different VM types. \FIG~\ref{fig:diss_vm_type} shows the startup time differences in Mao et al.'s measurements and our measurements. Due to the 9 years gap, we could not directly compare the VM startup time using the same instance types. Instead, we tried to compare the startup time with similar VM types. As shown in \FIG~\ref{fig:diss_vm_type}, except for {\tt micro} instance types, the newer instances used in this work show a significant reduction (53\% -- 67\%) in their startup times. This result also confirms that newer generation instances often show faster startup times than older generation instances, which is one of our findings described in Section~\ref{subsec:instance_type}.

\begin{figure}[t]
	\centering
		\includegraphics[width=1\columnwidth]{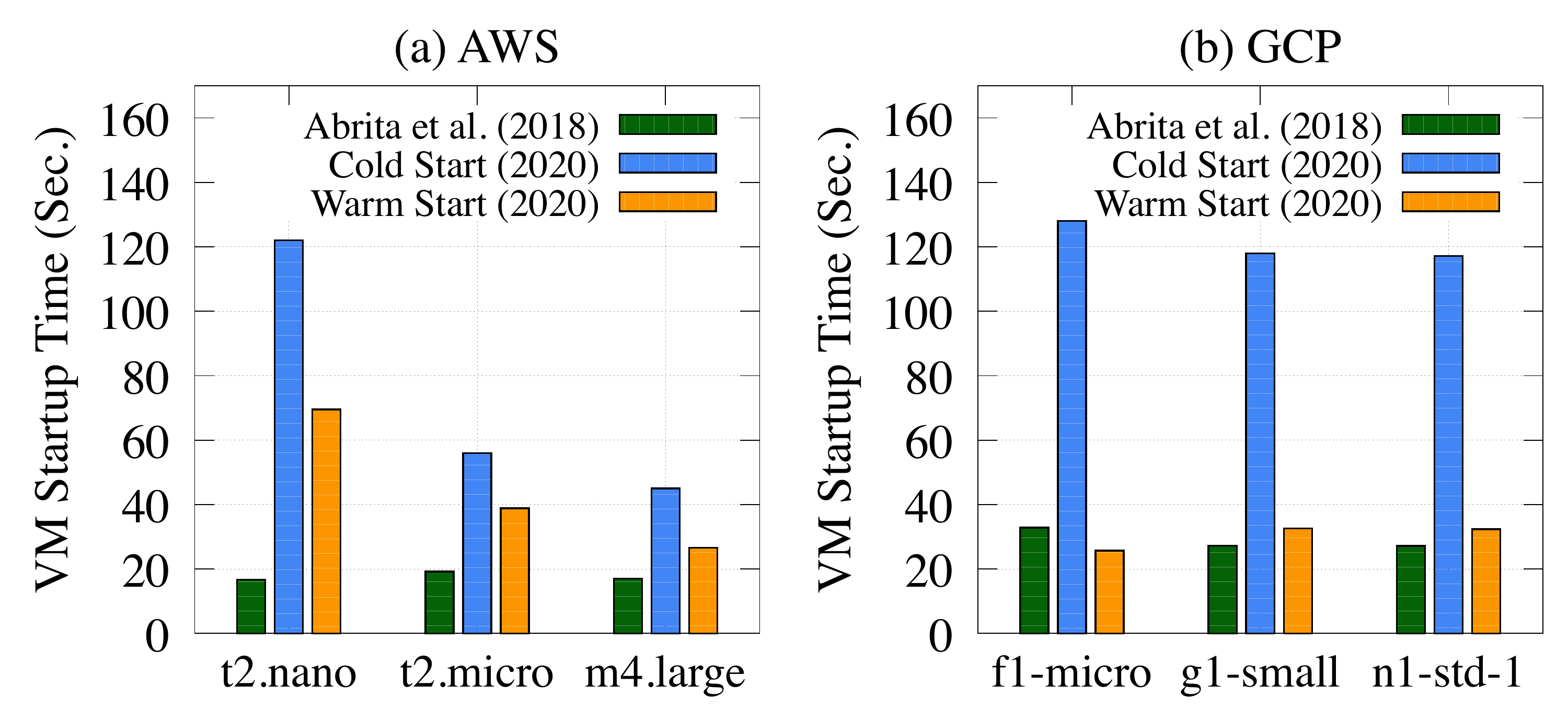}
		\caption{Comparison with previous work (Abrita et al.~
		\cite{Provisioning:Bench18}); VM startup times with different VM types in AWS and GCP. Note that the VM image size measured by Abrita et al. was not specified.}
		\label{fig:diss_vm_type_2018}
\end{figure}

\FIG~\ref{fig:diss_vm_type_2018} reports the comparison of VM startup times between the previous report by Abrita et al. (2018) and this measurement. 
For AWS, our work and Abrita et al. commonly measured startup times of three VM types; {\tt t2.nano}, {\tt t2.micro}, and {\tt m4.large}. 
The AWS VM startup times shown in \FIG~\ref{fig:diss_vm_type_2018}(a) are considerably different from the previous comparison with Mao et al. Specifically, Abrita et al. reported much faster VM startup times (17 -- 19 seconds) than our measurements, and their results are even faster than warm startup times of these three VM types. We believe this difference can be because Abrita et al. used different VM image sizes and/or templates than those used in our work. In the early stage of this work, we observed similar startup times when creating VMs with default OS images (without extra data in the image). 
Since detailed information regarding VM images used in Abrita et al. is unknown, we presume that using different VM images {\em may} be the reason for the faster startup time reported in Abrita et al.'s work.

\FIG~\ref{fig:diss_vm_type_2018}(b) shows the startup time comparisons in GCP against the measurement results by Abrita et al. We compared the startup time results from  {\tt f1-micro}, {\tt g1-small}, and {\tt n1-standard-1} VM types because these VM types were commonly measured. 
The results from Abrita et al. showed much faster VM startup times than our measurement of cold startup times, but Abrita et al's results are very similar to the warm startup times of these VM types. We tried to perform a further investigation to understand the differences/similarities, but \cite{Provisioning:Bench18} lacks the description of measurement methodology (especially for the procedure to measure cold/warm startup times). Therefore, we assume that the majority of the measurements by Abrita et al. contains the warm startup time results.

\subsubsection{VM Startup Time Comparison with Different Regions}
Our measurements for VM startup times per different regions are also compared with Abrita et al.'s benchmark results. In this comparison, the startup times of {\tt t2.micro} (AWS) in three regions and GCP {\tt f1-micro}'s startup times for three regions are used. Table~\ref{tab:comparison_region_islam18} reports the comparison results. For both providers, the VM startup times' differences are similar to the previous VM types comparisons reported in Section~\ref{subsubsec:compare_vmtype}. {\tt t2.micro} of AWS showed faster VM startup times in Abrita et al.'s report than warm startup times from our work. We also assume that this can be because of the VM image size differences and the use of default VM images. In GCP, {\tt f1-micro}'s startup times measured by Abrita et al. are also very close to warm startup times from our measurements; thus, we hold the same conclusion described in Section~\ref{subsubsec:compare_vmtype}.

\subsubsection{Spot Instance Startup Time Comparison} 
The spot VMs' startup times are compared with the results reported by Mao et al., and Table~\ref{tab:google_preemptive} reports the comparison results. As shown in the table, the spot instances in 2012 had 500 -- 595 seconds of startup times. A significant portion of such slow startup times was mainly related to the bidding process for determining the spot price~\cite{SpotPerf:CLOUD18, Spot:TOMPECS18}. However, after AWS released a new spot pricing model in 2018~\cite{AWS-new-spot-online}, such a slower bidding process has been replaced with a simplified mechanism so that AWS users can quickly move on to the VM startup process. For a fair comparison, we calculate VM startup time without a bidding process by subtracting bidding time from the whole spot VM startup time reported from the original author. The calculated results are shown in Table~\ref{tab:google_preemptive} (please refer to the values in the parentheses), and the startup times after the spot bidding process are from 95 to 160 seconds These results are similar to on-demand VM startup times reported in \FIG~\ref{fig:diss_vm_type}, except for {\tt t1.micro}, which is 58\% slower (about 60 seconds in on-demand VMs, 95s in spot VMs). However, our measurements show that spot instances have 41 to 53 seconds of startup times, which are 43\% -- 69\% quicker than the measurements in 2012, indicating that AWS improves spot infrastructure significantly and provides much faster startup times. 

\begin{table}[]
\centering
\caption{Comparison with previous work (Abrita et al.~\cite{Provisioning:Bench18}); VM startup times in different regions of AWS and GCP.}
\begin{tabular}{|l|l|l|l|l|}
\hline
\multirow{2}{*}{\textbf{\begin{tabular}[c]{@{}l@{}}VM Type\\ (Provider)\end{tabular}}} & \multirow{2}{*}{\textbf{Region}} & \multirow{2}{*}{\textbf{\begin{tabular}[c]{@{}l@{}}2018\\ Measure.\end{tabular}}} & \multicolumn{2}{l|}{\textbf{Current Measure.}} \\ \cline{4-5} 
 &  &  & \textbf{Cold} & \textbf{Warm} \\ \hline
\multirow{3}{*}{\begin{tabular}[c]{@{}c@{}}{\tt t2.micro}\\ (AWS)\end{tabular}} & {\tt us-east-1} & 27.6s & 58.8s & 39.3s \\ \cline{2-5} 
 & {\tt us-west-2} & 30.2s & 61.5s & 41.0s \\ \cline{2-5} 
 & {\tt ap-s.east-1} & 26.2s & 54.6s & 35.6s \\ \hline
\multirow{3}{*}{\begin{tabular}[c]{@{}c@{}}{\tt f1-micro}\\ (GCP)\end{tabular}} & {\tt eu-west1-c} & 33.0s & 116.9s & 33.7s \\ \cline{2-5} 
 & {\tt us-west1-a} & 36.1s & 133.9s & 35.1s \\ \cline{2-5} 
 & {\tt asia-s.east1-a} & 32.8s & 130.1s & 28.7s \\ \hline
\end{tabular}
\label{tab:comparison_region_islam18}
\end{table}

\vspace{1mm}
\noindent
\textbf{Summary of Comparison.} 
We compared our measurements with the two most relevant previous reports for VM startup times. 
By comparing with the results by Mao et al., we confirm that AWS made significant improvements for the VM startup times and currently have much quicker VM startup times in both on-demand and spot instances. However, compared with Abrita et al.'s results, we could not confirm the improvements or changes in VM startup times because we were not able to reproduce their results. This could be because of differences in VM configurations and measurement methodologies. i.e., cold vs. warm startup times, VM image sizes.

\subsection{Use Cases}
The accurate knowledge of VM startup time is critical for designing effective
predictive auto-scaling
policies~\cite{Ming-Autscale-SC11, NetflixScryer:online, PreAutos:CLOUD2011, SelfAdaptive-ICPE2013, PredEval:Cloud2016, CloudInsight:Cloud2018, TOMPECS18:AutoscalingWF, IPDPS20:Dynamo, TCC20:CloudInsight}. In predictive auto-scaling, new VMs are provisioned in advance to handle the increased workloads before the increased workload arrives. Without accurate knowledge of the VM
startup time, the newly-provisioned VMs may be created too early, leading to
idle VMs and wasted resources. Without this knowledge, the newly provisioned VMs
may also be started too late. The delayed provision makes resource under-provisioning, resulting in low performance and Quality-of-Service (QoS) requirement violations.

The accurate knowledge of VM startup time is also crucial for reliable cloud
simulation. Instead of actual executions in the cloud, cloud simulators
typically use execution time profiles of the VMs in their
simulations~\cite{CloudSim:SPE2011, JGC12:iCanCloud, JSC12:GreenCloud, PICS:CLOUD2015, TrustSim-IC2E15, ResMgmtTesting-ICPE18}. Therefore, the accuracy of the simulation results depends on the
accuracy of these profiles. Furthermore, as reported in this paper, the VM startup time
can be considerably longer and thus cannot be ignored. Therefore, accurate VM
startup time is also an important factor to be considered by cloud simulators
for reliable simulation results.

\section{Related Work}\label{sec:relatedwork}
There has been a large body of works that measured the performance of cloud infrastructures, such as CPU, IO, Network, and application performance~\cite{RuntimeMeasurement:VLDB10, PerfVar:TPDS11, Chaos:TIT2016, PerfVarWei:FSE19, EarlyObservation:HPDC10, MemPerf:TCC17, IOPerf:JGC13, NetPerf:CN15, AWSNetwork:Sigmetrics15, StartupMao:CLOUD12, Provisioning:Bench18, Less:CLOUD15}. Regarding the VM startup time, Mao et al.~\cite{StartupMao:CLOUD12} performed a measurement study on VM startup times from AWS EC2, Azure, and Rackspace in 2012. The authors identified several factors that could affect the VM startup time.
Abrita et al.~\cite{Provisioning:Bench18} reported benchmarking results of VM startup times in three cloud providers. 
However, both previous works' measurement results have limitations to be used for the current cloud research and resource management. 
First, the measurement from Mao et al. was performed 9 years ago and the results do not correctly reflect the current status of VM startup times. This is because a number of mechanisms~\cite{ScalabeVMDeployment:SC13, Squirrel:HPDC14, muVM:ICDCS15, Xu:JSS16, LigherVM-OSDI2017, CooporativeCaching:TCC2018, YOLO:Europar2019, Firecracker-NSDI2020} have been proposed and applied for optimizing VM provisioning processes in cloud data centers so that it is necessary to measure updated VM startup times to provide up-to-date information to the research community. Second, the report from Abrita et al. provides recent statistics in VM startup times, but the benchmarking results employed the limited VM configurations with unclear measurement methodologies. 
Moreover, our analysis and measurement study considered a much broader set of VM configurations (768 configurations in AWS, 704 configurations in GCP), including instance types, image sizes, and data center locations, and we conducted this measurement study for a much more extended period of time (three months). For example, Mao et al. used 6 VM types in AWS, and Abrita et al. only considered 5 VM types in AWS and 3 VM types in GCP, but our work employed more than 11 VM types from each provider. 
Our measurement results were collected across 12+ zones in four regions from both providers, and we were able to collect and analyze \datanum data points for VM startup times. Therefore, we report a much more comprehensive and thorough analysis of VM startup times with diverse configurations and realistic scenarios.

\begin{table}[]
\centering
\caption{Comparison with previous work; VM startup times of spot instances. For the 2012 measurement by Mao et al.~\cite{StartupMao:CLOUD12}, the numbers in parentheses are VM provisioning time after finishing bidding process of spot instances.}
\begin{tabular}{lll}
\hline
\multicolumn{1}{|l|}{} & \multicolumn{1}{l|}{\begin{tabular}[c]{@{}c@{}}{\bf 2012}\\ {\bf Measurements}\end{tabular}} & \multicolumn{1}{l|}{\begin{tabular}[c]{@{}c@{}}{\bf 2020}\\ {\bf Measurements}\end{tabular}} \\ \hline
\multicolumn{1}{|l|}{micro ({\tt t1} vs. {\tt t3})} & \multicolumn{1}{c|}{550s (95s)} & \multicolumn{1}{c|}{53.53s} \\ \hline
\multicolumn{1}{|l|}{small ({\tt m1} vs. {\tt t3})} & \multicolumn{1}{c|}{595s (100s)} & \multicolumn{1}{c|}{41.62s} \\ \hline
\multicolumn{1}{|l|}{medium ({\tt c1} vs. {\tt t3a})} & \multicolumn{1}{c|}{500s (95s)} & \multicolumn{1}{c|}{41.80s} \\ \hline
\multicolumn{1}{|l|}{large ({\tt m1} vs. {\tt m5a})} & \multicolumn{1}{c|}{590s (160s)} & \multicolumn{1}{c|}{42.98s} \\ \hline
\multicolumn{1}{|l|}{xlarge ({\tt m2} vs. {\tt m5})} & \multicolumn{1}{c|}{500s (120s)} & \multicolumn{1}{c|}{39.75s} \\ \hline
 &  & 
\end{tabular}
\label{tab:google_preemptive}
\end{table}

\section{Conclusion}\label{sec:conclusion}
In this work, we report the analysis of VM startup times in AWS and GCP. We measured and analyzed VM startup times with diverse configurations, which include two different OS types, 11+ VM types, four different VM image sizes (from 32GB to 256GB), four geographical regions (two in the U.S., one in Europe, and one in Asia), and two purchase models (on-demand vs. spot/preemptible). We collected more than \datanum data points from each provider with three months of measurements, analyzed these data points, and identified factors that change VM startup times. Our measurement results show that VM startup time can be varying significantly due to diverse factors such as OS types, VM image sizes, VM instance types, data center locations, and a caching mechanism in a cloud provider (e.g., GCP).

We then compared measurement and analysis results from this work against prior measurement studies. By comparing with the 2012 measurement~\cite{StartupMao:CLOUD12}, we confirmed that cloud providers (specifically AWS) made significant improvements for the VM startup times and currently offer much quicker VM startup times compared to the previous measurement results.

\section*{Acknowledgment}\label{sec:ack}
The authors thank Radhika Bhavsar for contributions to earlier versions of this work. 

\bibliographystyle{unsrt}
\bibliography{reference.bib}

\begin{thebibliography}{10}

\bibitem{ViewCloud:CACM10}
Michael Armbrust, Armando Fox, Rean Griffith, Anthony~D. Joseph, Randy~H. Katz,
  Andy Konwinski, Gunho Lee, David~A. Patterson, Ariel Rabkin, Ion Stoica, and
  Matei Zaharia.
\newblock {A View of Cloud Computing}.
\newblock {\em Communications of {ACM}}, 53(4):50--58, 2010.

\bibitem{AWS-online}
Amazon {W}eb {S}ervices.
\newblock \url{http://aws.amazon.com}, 2021.

\bibitem{GoogleCloud-online}
{G}oogle {C}loud {P}latform.
\newblock \url{https://cloud.google.com}, 2021.

\bibitem{OpenStack-online}
{O}pen{S}tack.
\newblock \url{https://www.openstack.org/}, 2021.

\bibitem{Docker-online}
{Docker}.
\newblock \url{https://www.docker.com/}, 2021.

\bibitem{gvisor-online}
{gVisor}.
\newblock \url{https://github.com/google/gvisor}, 2021.

\bibitem{MESOS:NSDI11}
Benjamin Hindman, Andy Konwinski, Matei Zaharia, Ali Ghodsi, Anthony~D. Joseph,
  Randy~H. Katz, Scott Shenker, and Ion Stoica.
\newblock {Mesos: A Platform for Fine-Grained Resource Sharing in the Data
  Center}.
\newblock In {\em {USENIX} Symposium on Networked Systems Design and
  Implementation (NSDI)}, Boston, MA, April, 2011.

\bibitem{Borg:Eurosys20}
Muhammad Tirmazi, Adam Barker, Nan Deng, Md~E. Haque, Zhijing~Gene Qin, Steven
  Hand, Mor Harchol{-}Balter, and John Wilkes.
\newblock {Borg: the Next Generation}.
\newblock In {\em European Conference on Computer Systems (EuroSys)},
  Heraklion, Greece, April, 2020.

\bibitem{AWSLambda-online}
{AWS Lambda}.
\newblock \url{https://aws.amazon.com/lambda/}, 2021.

\bibitem{Occupy:SoCC17}
Eric Jonas, Qifan Pu, Shivaram Venkataraman, Ion Stoica, and Benjamin Recht.
\newblock {Occupy the Cloud: Distributed Computing for the 99{\%}}.
\newblock In {\em ACM Symposium on Cloud Computing (SoCC)}, Santa Clara, CA,
  USA, September, 2017.

\bibitem{CACM19:RiseServerless}
Paul~C. Castro, Vatche Ishakian, Vinod Muthusamy, and Aleksander Slominski.
\newblock {The rise of serverless computing}.
\newblock {\em Communications of {ACM}}, 62(12):44--54, 2019.

\bibitem{ecs-online}
{Amazon Elastic Container Service}.
\newblock \url{https://aws.amazon.com/ecs/}, 2021.

\bibitem{k8s-online}
{Kubernetes}.
\newblock \url{https://kubernetes.io/}, 2021.

\bibitem{Catalyzer:ASPLOS2020}
Dong Du, Tianyi Yu, Yubin Xia, Binyu Zang, Guanglu Yan, Chenggang Qin, Qixuan
  Wu, and Haibo Chen.
\newblock {Catalyzer: Sub-millisecond Startup for Serverless Computing with
  Initialization-less Booting}.
\newblock In {\em ACM International Conference on Architectural Support for
  Programming Languages and Operating Systems (ASPLOS)}, Lausanne, Switzerland,
  March, 2020.

\bibitem{RuntimeMeasurement:VLDB10}
J{\"{o}}rg Schad, Jens Dittrich, and Jorge{-}Arnulfo Quian{\'{e}}{-}Ruiz.
\newblock {Runtime Measurements in the Cloud: Observing, Analyzing, and
  Reducing Variance}.
\newblock {\em Proc. {VLDB} Endow.}, 3(1):460--471, 2010.

\bibitem{PerfVar:TPDS11}
Alexandru Iosup, Nezih Yigitbasi, and Dick H.~J. Epema.
\newblock {On the Performance Variability of Production Cloud Services}.
\newblock In {\em {IEEE/ACM} International Symposium on Cluster, Cloud and Grid
  Computing (CCGrid)}, Newport Beach, CA, USA, May, 2011.

\bibitem{Chaos:TIT2016}
Philipp Leitner and J{\"{u}}rgen Cito.
\newblock {Patterns in the Chaos -- A Study of Performance Variation and
  Predictability in Public IaaS Clouds}.
\newblock {\em {ACM Trans. on Internet Technology}}, 16(3):15:1--15:23, 2016.

\bibitem{PerfVarWei:FSE19}
Sen He, Glenna Manns, John Saunders, Wei Wang, Lori~L. Pollock, and Mary~Lou
  Soffa.
\newblock {A Statistics-based Performance Testing Methodology for Cloud
  Applications}.
\newblock In {\em {ACM} Joint Meeting on European Software Engineering
  Conference and Symposium on the Foundations of Software Engineering
  (ESEC/FSE)}, Tallinn, Estonia, August, 2019.

\bibitem{EarlyObservation:HPDC10}
Zach Hill, Jie Li, Ming Mao, Arkaitz Ruiz{-}Alvarez, and Marty Humphrey.
\newblock {Early Observations on the Performance of Windows Azure}.
\newblock In {\em {ACM} International Symposium on High Performance Distributed
  Computing (HPDC)}, Chicago, Illinois, USA, June, 2010.

\bibitem{MemPerf:TCC17}
Iman Sadooghi, Jesus~Hernandez Martin, Tonglin Li, Kevin Brandstatter, Ketan
  Maheshwari, Tiago Pais Pitta De~Lacerda Ruivo, Gabriele Garzoglio, Steven
  Timm, Yong Zhao, and Ioan Raicu.
\newblock {Understanding the Performance and Potential of Cloud Computing for
  Scientific Applications}.
\newblock {\em {IEEE} Transactions on Cloud Computing}, 5(2):358--371, 2017.

\bibitem{IOPerf:JGC13}
Roberto~R. Exp{\'{o}}sito, Guillermo~L. Taboada, Sabela Ramos, Jorge
  Gonz{\'{a}}lez{-}Dom{\'{\i}}nguez, Juan Touri{\~{n}}o, and Ramon Doallo.
\newblock {Analysis of I/O Performance on an Amazon EC2 Cluster Compute and
  High I/O Platform}.
\newblock {\em Journal of Grid Computing}, 11(4):613--631, 2013.

\bibitem{NetPerf:CN15}
Valerio Persico, Pietro Marchetta, Alessio Botta, and Antonio Pescap{\`{e}}.
\newblock {Measuring network throughput in the cloud: The case of Amazon EC2}.
\newblock {\em Computer Networks}, 93:408--422, 2015.

\bibitem{Less:CLOUD15}
Jiawei Wen, Lei Lu, Giuliano Casale, and Evgenia Smirni.
\newblock {Less Can Be More: Micro-managing VMs in Amazon EC2}.
\newblock In {\em {IEEE} International Conference on Cloud Computing (CLOUD)},
  New York City, NY, USA, June, 2015.

\bibitem{AWSNetwork:Sigmetrics15}
Anshul Gandhi and Justin Chan.
\newblock {Analyzing the Network for {AWS} Distributed Cloud Computing}.
\newblock {\em {SIGMETRICS Performance Evaluation Review}}, 43(3):12--15, 2015.

\bibitem{StartupMao:CLOUD12}
Ming Mao and Marty Humphrey.
\newblock {A Performance Study on the VM Startup Time in the Cloud}.
\newblock In {\em {IEEE} International Conference on Cloud Computing (CLOUD)},
  Honolulu, HI, USA, June, 2012.

\bibitem{Provisioning:CCGrid14}
Gabriella Carrozza, Luigi Battaglia, Vittorio Manetti, Antonio Marotta, Roberto
  Canonico, and Stefano Avallone.
\newblock {On the Evaluation of VM Provisioning Time in Cloud Platforms for
  Mission-Critical Infrastructures}.
\newblock In {\em {IEEE/ACM} International Symposium on Cluster, Cloud and Grid
  Computing (CCGrid)}, Chicago, IL, USA, May, 2014.

\bibitem{Provisioning:Bench18}
Samiha~Islam Abrita, Moumita Sarker, Faheem Abrar, and Muhammad~Abdullah Adnan.
\newblock {Benchmarking {VM} Startup Time in the Cloud}.
\newblock In {\em Benchmarking, Measuring, and Optimizing - BenchCouncil
  International Symposium (Bench)}, Seattle, WA, USA, December, 2018.

\bibitem{Ming-Autscale-SC11}
Ming Mao and Marty Humphrey.
\newblock {Auto-scaling to minimize cost and meet application deadlines in
  cloud workflows}.
\newblock In {\em International Conference on High Performance Computing
  Networking, Storage and Analysis (SC)}, Seattle, WA, USA, November, 2011.

\bibitem{PredEval:Cloud2016}
In~Kee Kim, Wei Wang, Yanjun Qi, and Marty Humphrey.
\newblock {Empirical Evaluation of Workload Forecasting Techniques for
  Predictive Cloud Resource Scaling}.
\newblock In {\em {IEEE} International Conference on Cloud Computing (CLOUD)},
  San Francisco, CA, USA, June, 2016.

\bibitem{ScalabeVMDeployment:SC13}
Kaveh Razavi and Thilo Kielmann.
\newblock {Scalable Virtual Machine Deployment Using VM Image Caches}.
\newblock In {\em International Conference for High Performance Computing,
  Networking, Storage and Analysis (SC)}, Denver, CO, {USA}, November, 2013.

\bibitem{Squirrel:HPDC14}
Kaveh Razavi, Ana Ion, and Thilo Kielmann.
\newblock {Squirrel: scatter hoarding {VM} image contents on IaaS compute
  nodes}.
\newblock In {\em International Symposium on High-Performance Parallel and
  Distributed Computing (HPDC)}, Vancouver, BC, Canada, June, 014.

\bibitem{muVM:ICDCS15}
Kaveh Razavi, Gerrit Van~Der Kolk, and Thilo Kielmann.
\newblock {Prebaked {\(\mathrm{\mu}\)}VMs: Scalable, Instant {VM} Startup for
  IaaS Clouds}.
\newblock In {\em {IEEE} International Conference on Distributed Computing
  Systems (ICDCS)}, Columbus, OH, June, 2015.

\bibitem{Xu:JSS16}
Jiwei Xu, Wenbo Zhang, Zhenyu Zhang, Tao Wang, and Tao Huang.
\newblock Clustering-based acceleration for virtual machine image deduplication
  in the cloud environment.
\newblock {\em Journal of Systems and Software}, 121:144--156, 2016.

\bibitem{LigherVM-OSDI2017}
Filipe Manco, Costin Lupu, Florian Schmidt, Jose Mendes, Simon Kuenzer, Sumit
  Sati, Kenichi Yasukata, Costin Raiciu, and Felipe Huici.
\newblock {My {VM} is Lighter (and Safer) than your Container}.
\newblock In {\em ACM Symp. on Operating Systems Principles (SOSP)}, Shanghai,
  China, October, 2017.

\bibitem{CooporativeCaching:TCC2018}
Yifan Zhang, Kai Niu, Weigang Wu, Keqin Li, and Yu~Zhou.
\newblock {Speeding Up VM Startup by Cooperative VM Image Caching}.
\newblock {\em IEEE Transactions on Cloud Computing}, pages 1--1, 2018.

\bibitem{YOLO:Europar2019}
Thuy{-}Linh Nguyen, Ramon Nou, and Adrien Lebre.
\newblock {YOLO: Speeding Up VM and Docker Boot Time by Reducing I/O
  Operations}.
\newblock In {\em European Conference on Parallel and Distributed Processing
  (Euro-Par)}, G{\"{o}}ttingen, Germany, August, 2019.

\bibitem{Firecracker-NSDI2020}
Alexandru Agache, Marc Brooker, Alexandra Iordache, Anthony Liguori, Rolf
  Neugebauer, Phil Piwonka, and Diana{-}Maria Popa.
\newblock {Firecracker: Lightweight Virtualization for Serverless
  Applications}.
\newblock In {\em {USENIX} Symposium on Networked Systems Design and
  Implementation (NSDI)}, Santa Clara, CA, USA, February, 2020.

\bibitem{TPDS11:Iosup}
Alexandru Iosup, Simon Ostermann, Nezih Yigitbasi, Radu Prodan, Thomas
  Fahringer, and Dick H.~J. Epema.
\newblock {Performance Analysis of Cloud Computing Services for Many-Tasks
  Scientific Computing}.
\newblock {\em {IEEE} Transactions on Parallel and Distributed Systems},
  22(6):931--945, 2011.

\bibitem{UCC14:LCA}
In~Kee Kim, Jacob Steele, Yanjun Qi, and Marty Humphrey.
\newblock {Comprehensive Elastic Resource Management to Ensure Predictable
  Performance for Scientific Applications on Public IaaS Clouds}.
\newblock In {\em {IEEE/ACM} International Conference on Utility and Cloud
  Computing (UCC)}, London, United Kingdom, December, 2014.

\bibitem{TOMPECS18:AutoscalingWF}
Alexey Ilyushkin, Ahmed Ali{-}Eldin, Nikolas Herbst, Andr{\'{e}} Bauer,
  Alessandro~Vittorio Papadopoulos, Dick H.~J. Epema, and Alexandru Iosup.
\newblock {An Experimental Performance Evaluation of Autoscalers for Complex
  Workflows}.
\newblock {\em ACM Transactions on Modeling and Performance Evaluation of
  Computing Systems}, 3(2):8:1--8:32, 2018.

\bibitem{CloudInsight:Cloud2018}
In~Kee Kim, Wei Wang, Yanjun Qi, and Marty Humphrey.
\newblock {CloudInsight: Utilizing a Council of Experts to Predict Future Cloud
  Application Workloads}.
\newblock In {\em IEEE International Conference on Cloud Computing (CLOUD)},
  San Francisco, CA, USA, July, 2018.

\bibitem{IPDPS20:Dynamo}
Vinodh~Kumaran Jayakumar, Jaewoo Lee, In~Kee Kim, and Wei Wang.
\newblock {A Self-Optimized Generic Workload Prediction Framework for Cloud
  Computing}.
\newblock In {\em {IEEE} International Parallel and Distributed Processing
  Symposium (IPDPS)}, Virtual Event, May, 2020.

\bibitem{TCC20:CloudInsight}
In~Kee Kim, Wei Wang, Yanjun Qi, and Marty Humphrey.
\newblock {Forecasting Cloud Application Workloads with CloudInsight for
  Predictive Resource Management}.
\newblock {\em IEEE Transactions on Cloud Computing}, pages 1--1, 2020.

\bibitem{CloudSim:SPE2011}
Rodrigo~N. Calheiros, Rajiv Ranjan, Anton Beloglazov, C{\'{e}}sar A. F.~De
  Rose, and Rajkumar Buyya.
\newblock {CloudSim: a toolkit for modeling and simulation of cloud computing
  environments and evaluation of resource provisioning algorithms}.
\newblock {\em Software: Practice and Experience (SPE)}, 41(1):23--50, 2011.

\bibitem{JGC12:iCanCloud}
Alberto Nu{\~{n}}ez, Jos{\'{e}}~Luis V{\'{a}}zquez{-}Poletti, Agust{\'{\i}}n~C.
  Caminero, Gabriel~G. Casta{\~{n}}{\'{e}}, Jes{\'{u}}s Carretero, and
  Ignacio~Mart{\'{\i}}n Llorente.
\newblock {iCanCloud: {A} Flexible and Scalable Cloud Infrastructure
  Simulator}.
\newblock {\em Journal of Grid Computing}, 10(1):185--209, 2012.

\bibitem{JSC12:GreenCloud}
Dzmitry Kliazovich, Pascal Bouvry, and Samee~Ullah Khan.
\newblock {GreenCloud: a packet-level simulator of energy-aware cloud computing
  data centers}.
\newblock {\em Journal of Supercomputing}, 62(3):1263--1283, 2012.

\bibitem{PICS:CLOUD2015}
In~Kee Kim, Wei Wang, and Marty Humphrey.
\newblock {PICS: A Public IaaS Cloud Simulator}.
\newblock In {\em {IEEE} International Conference on Cloud Computing (CLOUD)},
  New York City, NY, USA, June, 2015.

\bibitem{TrustSim-IC2E15}
Alexander Pucher, Emre Gul, Rich Wolski, and Chandra Krintz.
\newblock {Using Trustworthy Simulation to Engineer Cloud Schedulers}.
\newblock In {\em {IEEE} International Conference on Cloud Engineering (IC2E)},
  Tempe, AZ, USA, March, 2015.

\bibitem{CLOUD2021:VMStartup}
Jianwei Hao, Ting Jiang, Wei Wang, and In~Kee Kim.
\newblock {An Empirical Analysis of VM Startup Times in Public IaaS Clouds}.
\newblock In {\em {IEEE} International Conference on Cloud Computing (CLOUD)},
  Virtual Event, September, 2021.

\bibitem{AWS-Spot-online}
{Amazon EC2 Spot Instances}.
\newblock \url{https://aws.amazon.com/ec2/spot/}, 2021.

\bibitem{GoogleCloud-PVM-online}
{Preemptible VM instances}.
\newblock \url{https://cloud.google.com/compute/docs/instances/preemptible},
  2021.

\bibitem{MonDelay:FGCS2013}
Michael Smit, Bradley Simmons, and Marin Litoiu.
\newblock Distributed, application-level monitoring for heterogeneous clouds
  using stream processing.
\newblock {\em Future Generation Computing Systems}, 29(8):2103--2114, 2013.

\bibitem{AWS-boto-online}
{AWS SDK for Python (Boto3)}.
\newblock \url{https://aws.amazon.com/sdk-for-python/}, 2021.

\bibitem{GoogleCloud-API-online}
{Compute Engine client libraries}.
\newblock
  \url{https://cloud.google.com/compute/docs/api/libraries\#google_apis_python_client_library},
  2021.

\bibitem{Taming:OSDI2018}
Aleksander Maricq, Dmitry Duplyakin, Ivo Jimenez, Carlos Maltzahn, Ryan
  Stutsman, Robert Ricci, and Ana Klimovic.
\newblock {Taming Performance Variability}.
\newblock In {\em {USENIX} Symposium on Operating Systems Design and
  Implementation (OSDI)}, Carlsbad, CA, USA, October, 2018.

\bibitem{VMCaching:Europar2012}
Pradipta De, Manish Gupta, Manoj Soni, and Aditya Thatte.
\newblock {Caching VM Instances for Fast VM Provisioning: A Comparative
  Evaluation}.
\newblock In {\em European Conference on Parallel Processing (Euro-Par)},
  Rhodes Island, Greece, August, 2012.

\bibitem{GoogleCloud-MachineTypes}
{Google Compute Engine Machine types}.
\newblock \url{https://cloud.google.com/compute/docs/machine-types}, 2021.

\bibitem{ZillowSpot:online}
Zillow.
\newblock {Saving Money with EMR Auto Scaling and Spot Instances}.
\newblock \url{https://www.zillow.com/tech/save-money-emr-autoscaling-spot/},
  2017.

\bibitem{SpotPerf:CLOUD18}
Thanh{-}Phuong Pham, Sasko Ristov, and Thomas Fahringer.
\newblock {Performance and Behavior Characterization of Amazon EC2 Spot
  Instances}.
\newblock In {\em IEEE International Conference on Cloud Computing (CLOUD)},
  San Francisco, CA, USA, July, 2018.

\bibitem{Spot:TOMPECS18}
Cheng Wang, Qianlin Liang, and Bhuvan Urgaonkar.
\newblock {An Empirical Analysis of Amazon EC2 Spot Instance Features Affecting
  Cost-Effective Resource Procurement}.
\newblock {\em ACM Transactions on Modeling and Performance Evaluation of
  Computing Systems}, 3(2):6:1--6:24, 2018.

\bibitem{AWS-new-spot-online}
Amazon~Web Services.
\newblock {New Amazon EC2 Spot pricing model: Simplified purchasing without
  bidding and fewer interruptions}.
\newblock
  \url{https://aws.amazon.com/blogs/compute/new-amazon-ec2-spot-pricing/},
  2018.

\bibitem{NetflixScryer:online}
Netflix~Technology Blog.
\newblock Scryer: Netflix’s predictive auto scaling engine.
\newblock
  \url{https://netflixtechblog.com/scryer-netflixs-predictive-auto-scaling-engine-a3f8fc922270},
  2013.

\bibitem{PreAutos:CLOUD2011}
N.~{Roy}, A.~{Dubey}, and A.~{Gokhale}.
\newblock {Efficient Autoscaling in the Cloud Using Predictive Models for
  Workload Forecasting}.
\newblock In {\em IEEE International Conference on Cloud Computing}, 2011.

\bibitem{SelfAdaptive-ICPE2013}
Nikolas~Roman Herbst, Nikolaus Huber, Samuel Kounev, and Erich Amrehn.
\newblock {Self-adaptive workload classification and forecasting for proactive
  resource provisioning}.
\newblock In {\em {ACM/SPEC} International Conference on Performance
  Engineering (ICPE)}, 2013.

\bibitem{ResMgmtTesting-ICPE18}
Christian Stier, J{\"{o}}rg Domaschka, Anne Koziolek, Sebastian Krach, Jakub
  Krzywda, and Ralf~H. Reussner.
\newblock {Rapid Testing of IaaS Resource Management Algorithms via Cloud
  Middleware Simulation}.
\newblock In {\em {ACM/SPEC} International Conference on Performance
  Engineering (ICPE)}, 2018.

\end{thebibliography}

\end{document}